# Machine-Learning-Based Multiple Abnormality Prediction with Large-Scale Chest Computed Tomography Volumes

Rachel Lea Draelos[a,b], David Dov[c], Maciej A. Mazurowski[c,d,e], Joseph Y. Lo[c,d,f], Ricardo Henao[c,e], Geoffrey D. Rubin[d], Lawrence Carin[a,c,g]

[a]Computer Science Department, Duke University, LSRC Building D101, 308 Research Drive, Duke Box 90129, Durham, North Carolina 27708-0129, United States of America
[b]School of Medicine, Duke University, DUMC 3710, Durham, North Carolina 27710, United States of America
[c]Electrical and Computer Engineering Department, Edmund T. Pratt Jr. School of Engineering, Duke University, Box 90291, Durham, North Carolina 27708, United States of America
[d]Radiology Department, Duke University, Box 3808 DUMC, Durham, North Carolina 27710, United States of America
[e]Biostatistics and Bioinformatics Department, Duke University, DUMC 2424 Erwin Road, Suite 1102 Hock Plaza, Box 2721 Durham, North Carolina 27710, United States of America
[f]Biomedical Engineering Department, Edmund T. Pratt Jr. School of Engineering, Duke University, Room 1427, Fitzpatrick Center (FCIEMAS), 101 Science Drive, Campus Box 90281, Durham, North Carolina 27708-0281, United States of America
[g]Statistical Science Department, Duke University, Box 90251, Durham, North Carolina 27708-0251, United States of America

Address for correspondence:
Rachel Lea Draelos
Email: rlb61@duke.edu
Department of Computer Science
308 Research Drive
Durham, North Carolina 27708-0129
United States

## *Abstract*

Machine learning models for radiology benefit from large-scale data sets with high quality labels for abnormalities. We curated and analyzed a chest computed tomography (CT) data set of 36,316 volumes from 19,993 unique patients. This is the largest multiply-annotated volumetric medical imaging data set reported. To annotate this data set, we developed a rule-based method for automatically extracting abnormality labels from free-text radiology reports with an average F-score of 0.976 (min 0.941, max 1.0). We also developed a model for multi-organ, multi-disease classification of chest CT volumes that uses a deep convolutional neural network (CNN). This model reached a classification performance of AUROC >0.90 for 18 abnormalities, with an average AUROC of 0.773 for all 83 abnormalities, demonstrating the feasibility of learning from unfiltered whole volume CT data. We show that training on more labels improves performance significantly: for a subset of 9 labels – nodule, opacity, atelectasis, pleural effusion, consolidation, mass, pericardial effusion, cardiomegaly, and pneumothorax – the model's average AUROC increased by 10% when the number of training labels was increased from 9 to all 83. All code for volume preprocessing, automated label extraction, and the volume abnormality prediction model is publicly available. The 36,316 CT volumes and labels will also be made publicly available pending institutional approval.

# *1 Introduction*

Automated interpretation of medical images using machine learning holds immense promise (Hosny et al., 2018; Kawooya, 2012; Schier, 2018). Machine learning models learn from data without being explicitly programmed and have demonstrated excellent performance across a variety of image interpretation tasks (Voulodimos et al., 2018). Possible applications of such models in radiology include human-computer interaction systems intended to further reduce the 3 – 5% real-time diagnostic error rate of radiologists (Lee et al., 2013) or automated triage systems that prioritize scans with urgent findings for earlier human assessment (Annarumma et al., 2019; Yates et al., 2018). Previous work applying machine learning to CT interpretation has focused on prediction of one abnormality at a time. Even when successful, such focused models have limited clinical applicability because radiologists are responsible for a multitude of findings in the images. To address this need, we investigate the simultaneous prediction of multiple abnormalities using a single model.

There has been substantial prior work on multiple-abnormality prediction in 2D projectional chest radiographs facilitated by the publicly available ChestX-ray14 (Wang et al., 2017), CheXpert (Irvin et al., 2019), and MIMIC-CXR (Johnson et al., 2019) datasets annotated with 14 abnormality labels. However, to the best of our knowledge, multilabel classification of whole 3D chest computed tomography (CT) volumes for a diverse range of abnormalities has not yet been reported. Prior work on CTs includes numerous models that evaluate one class of abnormalities at a time – *e.g.,* lung nodules (Ardila et al., 2019; Armato et al., 2011; Pehrson et al., 2019; Shaukat et al., 2019; Zhang et al., 2018), pneumothorax (Li et al., 2019), emphysema (Humphries et al., 2019), interstitial lung disease (Anthimopoulos et al., 2016; Bermejo-Peláez et al., 2020; Christe et al., 2019; Christodoulidis et al., 2017; Depeursinge et al., 2012; Gao et al., 2018, 2016; Walsh et al., 2018; Wang et al., 2019), liver fibrosis (Choi et al., 2018), colon polyps (Nguyen et al., 2012), renal cancer (Linguraru et al., 2011), vertebral fractures (Burns et al., 2016), and intracranial hemorrhage (Kuo et al., 2019; Lee et al., 2019). The public DeepLesion dataset (Yan et al., 2018) has enabled multiple studies on detection of focal lesions (Khajuria et al., 2019; Shao et al., 2019). There are three obstacles to large-scale multilabel classification of whole CTs: acquiring sufficently large datasets, preparing labels for each volume, and the technical challenges of developing a large-scale multi-label machine learning model for the task. In this study, we address all of these challenges in order to present a fully automated algorithm for multi-organ and multi-disease diagnosis in chest CT.

Acquiring a large CT dataset appropriate for computational analysis is challenging. There is no standardized software for bulk downloading and preprocessing of CTs for machine learning purposes. Each CT scan is associated with multiple image sets ("series"), each comprising on the order of 100,000,000 voxels. These volumes need to be organized and undergo many pre-processing steps.

To train a multilabel classification model, each volume must be associated with structured labels indicating the presence or absence of abnormalities. Given the number of organs and diseases, manual abnormality labeling by radiologists for the thousands of cases required to train an accurate machine learning model is virtually impossible. Instead, methods that automatically extract accurate labels from radiology reports are necessary (Irvin et al., 2019; Johnson et al., 2019; Wang et al., 2017).

Prior work in automated label extraction from radiology reports can be divided into two primary categories: whole-report classifiers (Banerjee et al., 2017; Chen et al., 2018; Pham et al., 2014; Zech et al., 2018) that predict all labels of interest simultaneously from a numerical representation of the full text, and rule-based methods that rely on handcrafted rules to assign abnormality labels. Whole-report classifiers suffer two key drawbacks: they are typically uninterpretable and they require expensive, time-consuming manual labeling of training reports, where the number of manual labels scales linearly with the number of training reports and with the number of abnormalities. Rule-based systems (Chapman et al., 2001; Demner-

---

[1] Abbreviations: SARLE (Sentence Analysis for Radiology Label Extraction), RAD-ChestCT (Report-Annotated Duke Chest CT), AUROC (area under the receiver operating characteristic)

Fushman et al., 2016; Irvin et al., 2019; Peng et al., 2018) are a surprisingly good alternative, as radiology language is rigid in subject matter, content, and spelling. We propose and validate a rule-based label extraction approach for chest CT reports designed to extract 83 abnormality labels.

Development of a multi-label classification model is challenging due to the complexity of multi-organ, multi-disease identification from CT scans. We will show that the frequency of particular abnormalities in CTs varies greatly, from nodules (78%) to hemothorax (<1%). There are hundreds of possible abnormalities; multiple abnormalities usually occur in the same scan (10±6); and the same abnormality can occur in multiple locations in one scan. Different abnormalities can appear visually similar, *e.g.,* atelectasis and pneumonia (Edwards et al., 2016), and the same abnormality can look visually different depending on severity (*e.g.,* pneumonia of one lobe *vs.* an entire lung) (Franquet, 2001), shape (*e.g.,* smooth nodule *vs.* spiculated nodule), and texture (*e.g.,* reticular *vs.* groundglass) (Dhara et al., 2016). Variation in itself is not necessarily pathologic – even among "normal" scans the body's appearance differs based on age, gender, weight, and natural anatomical variants (Hansell, 2010; Terpenning and White, 2015). Furthermore, there are hardly any "normal" scans available to teach the model what "normality" is. We will show that <1% of chest CTs in our data are "normal" (*i.e.,* lacking any of the 83 considered abnormalities). This low rate of normality is likely a reflection of requisite pre-test probabilities for disease that physicians consider before recommending CT and its associated exposure to ionizing radiation (Costello et al., 2013; Purysko et al., 2016; Smith-Bindman et al., 2009).

Previous single-abnormality CT classification studies have relied on time-intensive manual labeling of CT pixels (Kuo et al., 2019; Li et al., 2019; Walsh et al., 2018), patches (Anthimopoulos et al., 2016; Bermejo-Peláez et al., 2020; Christodoulidis et al., 2017; Gao et al., 2018), or slices (Gao et al., 2016; Lee et al., 2019) that typically limits the size of the data set to <1,000 CTs and restricts the total number of abnormalities that can be considered. Inspired by prior successes in the field of computer vision on identifying hundreds of classes in whole natural images (Deng et al., 2009; Rawat and Wang, 2017), we hypothesize that it should be possible to learn multi-organ, multi-disease diagnosis from whole CT data given sufficient training examples. We build a model that learns directly from whole CT volumes without any pixel, patch, or slice-level labels, and find that transfer learning and aggregation of features across the craniocaudal extent of the scan enables high performance on numerous abnormalities.

In this study we address the challenges of CT data preparation, automated label extraction from free-text radiology reports, and simultaneous multiple abnormality prediction from CT volumes using a deep convolutional neural network. We hope that this work will contribute to the long-term goal of automated radiology systems that assist radiologists, accelerate the medical workflow, and benefit patient care.

## 2 Methods

An overview of this study is shown in Figure 1.

**2.1 Chest CT Data Set Preparation**

The preparation of our retrospective chest CT data set included four stages: report download, report processing, volume download, and volume processing. The final dataset of 36,316 chest CT volumes obtained without intravenous contrast material and their associated reports from Duke University Health System spans January 2012 – April 2017 and was collected under IRB approval and in compliance with HIPAA. Informed consent was waived by the IRB.

In the first stage, 440,822 CT reports were obtained using the electronic health record and the Duke Enterprise Data Unified Content Explorer (DEDUCE) (Horvath et al., 2011) search tool. After filtering to remove duplicates, preliminary reports, un-addended versions, and empty reports, a dataset of 336,800 unique CT reports was obtained. This data set included head, chest, abdomen, and pelvis CTs, and include both intravenous contrast material enhanced and unenhanced scans. We then selected the 11% of reports (36,861) of chest CTs performed without intravenous contrast material based on the "protocol description" field.

Report text was prepared with standard natural language processing (NLP) preprocessing steps including lowercasing, replacing all whitespace with a single space, and removal of punctuation except for the periods inside decimal numbers which carry medical meaning (*e.g.,* 1.2 cm mass versus 12 cm mass). We replaced all times with a "%time" token, dates with "%date", and years with "%year."

The chest CT scans were queried and downloaded using an application programming interface (API) developed for the Duke vendor neutral archive. All scans were stored and preprocessed within the Duke Protected Analytics Computing Environment (PACE) which is a secure virtual network space that enables approved users to work with identifiable protected health information.

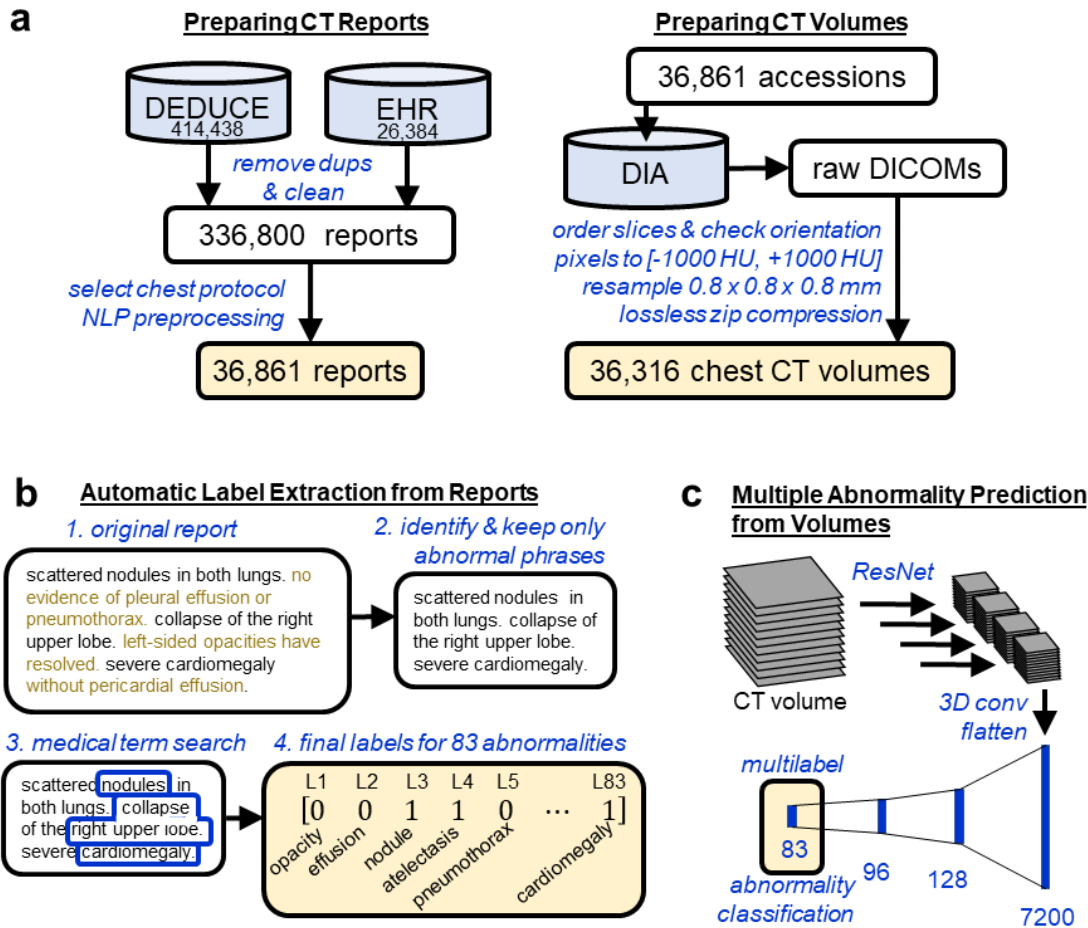

**Figure 1.** Study Overview. (a) Reports from chest CT scans performed without intravenous contrast material were acquired from the Duke Enterprise Data Unified Content Explorer (DEDUCE) search tool as well as the Epic electronic health record (EHR). Report accession numbers were used to download CT slices as DICOMs from the Duke Image Archive (DIA), which were processed into a final data set of 36,316 CT volumes. (b) We develop an approach for extracting binary labels for 83 different abnormalities from the free-text chest CT reports. (c) We train and evaluate a deep convolutional neural network model (shown here and detailed further in Figure 2) that takes as input a whole CT volume and predicts all 83 abnormality labels simultaneously.

Using DICOM header information, the original series with the most slices was selected for subsequent analysis, thus rejecting secondary, derived, or reformatted series. In total 36,316 volumes were acquired out of 36,861 initially specified. Across the 36,316 scans, the slice thickness and reconstruction intervals were 0.625 mm for 54% of scans, 0.6 mm for 41% of scans, and other values ≤5 mm for the remaining 5% of scans. The scans came from two vendors, General Electric (57%) and Siemens (43%), and twelve different CT scanner models, representing the entire fleet across the Duke Health system. 53% of scans were reconstructed using filtered back projection (FBP), while the remaining 47% were produced with iterative reconstruction. Common scan indications include pulmonary nodules, cancer, and interstitial lung disease.

An end-to-end Python pipeline was developed to process the separate DICOM files corresponding to different slices of one CT into a single 3D numpy array (Van Der Walt et al., 2011) compatible with the major machine learning frameworks PyTorch (Paszke et al., 2019) and TensorFlow (Abadi et al., 2016). CT sections were ordered and verified to be in a consistent orientation to facilitate future work in abnormality localization. Raw pixel values in DICOMs have undergone a linear transformation to enable efficient disk storage; this transformation was reversed to obtain pixel values in Hounsfield units (HU), using the DICOM attributes *RescaleSlope* and *RescaleIntercept*. Pixel values were clipped to [-1000 HU,+1000 HU], which represent practical lower and upper limits of the HU scale, corresponding to the radiodensities of air and dense bone respectively (DenOtter and Schubert, 2019; Lamba et al., 2014). Each volume was resampled using SimpleITK (Lowekamp et al., 2013) to $0.8 \times 0.8 \times 0.8$ mm to enable a consistent physical distance meaning of one pixel across all patients. To reduce storage requirements, the final 3D array was saved using lossless zip compression. The raw unprocessed DICOMs require 9.2 terabytes of storage. The final preprocessed arrays require 2.8 terabytes.

The data were randomly split into 70% volume training (25,355 volumes), 6% volume validation (2,085 volumes), 4% reserved for future studies (1,667), and 20% volume test (7,209 volumes) based on patient MRN so that no patient appears in more than one set. A subset of the volume training data was designated "report train" (639 reports) and "report test" (427 reports) and used to develop the automated label extraction method. Finally, we define a random subset of 2,000 training and 1,000 validation set scans that we use for architecture and ablation studies.

In total, 36,316 volumes paired with reports were successfully downloaded and prepared for analysis. We refer to the full data set as the Report-Annotated Duke Chest CTs (RAD-ChestCT) data set.

## 2.2 Automated image labeling through analysis of radiology reports

Manually recording the presence or absence of 83 different abnormalities for each of 36,316 volumes would require hand-coding over 3 million labels and is prohibitively time-consuming. To create a large data set of labeled CT volumes it is necessary to leverage automated label extraction approaches.

We follow the experimental setup of ChestX-Ray8 (Wang et al., 2017), CheXpert (Irvin et al., 2019), and MIMIC-CXR (Johnson et al., 2019), in which subsets of ~200 to 1,000 reports are used to develop and evaluate an automated label extraction method, and subsequently the final label extraction model is applied to the remaining tens of thousands of reports to obtain predicted labels. The predicted labels are then treated as ground truth in image-based experiments. We use 639 reports as a training set and 427 reports as a held-out test set.

The goal of radiology report label extraction is to analyze a free-text report and produce a binary vector of labels in which an entry is equal to one if the corresponding abnormality is present and is equal to zero otherwise. For example, if the predefined label order is [nodule, atelectasis, cardiomegaly] then the label [1,1,0] will be produced for a report in which nodule and atelectasis are present and cardiomegaly is absent.

A key challenge in radiology report label extraction is that the presence of a word, *e.g.*, "nodule," does not necessarily indicate presence of the abnormality in the scan, *e.g.*, "the previously seen nodule is no longer visualized." Furthermore, there are numerous phrasings for the same label, *e.g.*, "enlarged heart," "hypertrophic ventricles," and "severe cardiomegaly" for the label "cardiomegaly."

We propose two closely related approaches to extract 83 abnormalities labels from radiology reports, which we term Sentence Analysis for Radiology Label Extraction (SARLE). The first approach, SARLE-Hybrid, uses a machine learning sentence classifier followed by a rule-based term search, and is intended to be easier to adapt to text reports from other radiology modalities as it abstracts away abnormality detection rules. The second approach, SARLE-Rules, is fully rule-based and customized for chest CT reports. Both approaches are designed to be conceptually simple, to scale to a large number of abnormality labels, and to assign multiple abnormalities per scan when the report indicates that multiple abnormalities are present.

The language used in radiology reports varies by modality, anatomical region, and institution. We chose to create a custom lexicon targeted to the RAD-ChestCT dataset, rather than conforming a prospectively derived ontology to the dataset. Our 83 labels were chosen in an iterative manner with the goal of identifying all abnormalities mentioned in this dataset with at least ~2% frequency. First, we prepared an initial list of common abnormalities (*e.g.* nodule, cardiomegaly). Next, training set sentences that did not contain any of these abnormalities were examined in order to identify additional abnormalities. This process of abnormality sentence tagging followed by examination of untagged sentences was repeated until the final list of 83 abnormalities was obtained. The 83 extracted labels are shown in Table 1.

**Table 1. List of 83 Abnormalities that SARLE Extracts from Radiology Reports.** Note that each abnormality is associated with a set of medical synonyms that are defined in a term search step. For example, the term search for cardiomegaly captures "cardiomegaly," "dilated ventricles," "enlarged right atrium," and other synonyms. "Lung resection" captures pneumonectomy and lobectomy; "breast surgery" captures mastectomy and lumpectomy; pleural effusion captures "pleural effusion" and "pleural fluid" and so on. The term search for all abnormalities is available in Appendix B.

| | |
|---|---|
| Lung (22) | airspace disease, air trapping, aspiration, atelectasis, bronchial wall thickening, bronchiectasis, bronchiolectasis, bronchiolitis, bronchitis, consolidation, emphysema, hemothorax, interstitial lung disease, lung resection, mucous plugging, pleural effusion, infiltrate, pleural thickening, pneumonia, pneumonitis, pneumothorax, pulmonary edema, scattered nodules, septal thickening, tuberculosis |
| Lung Patterns (5) | bandlike or linear, groundglass, honeycombing, reticulation, tree in bud |
| Additional (47) | arthritis, atherosclerosis, aneurysm, breast implant, breast surgery, calcification, cancer, catheter or port, cavitation, clip, congestion, cyst, debris, deformity, density, dilation or ectasia, distention, fibrosis, fracture, granuloma, hardware, hernia, infection, inflammation, lesion, lucency, lymphadenopathy, mass, nodule, nodule >1 cm, opacity, plaque, postsurgical, scarring, scattered calcifications, secretion, soft tissue, staple, stent, suture, transplant, chest tube, tracheal tube, GI tube (includes NG and GJ tubes) |
| Heart (9) | CABG (coronary artery bypass graft), cardiomegaly, coronary artery disease, heart failure, heart valve replacement, pacemaker or defibrillator, pericardial effusion, pericardial thickening, sternotomy |

The first approach, SARLE-Hybrid, is a hybrid machine learning and rule-based method. The motivation behind this approach is to use machine learning to eliminate the need for hand-crafted negation detection rules found in fully rule-based methods (Chapman et al., 2001; Peng et al., 2018), while enabling scaling to a large number of abnormality labels through a rule-based term search that leverages medical vocabulary. Note that wholly machine learning approaches, *i.e.,* report classifiers (Chen et al., 2018; Pham et al., 2014; Zech et al., 2018) that take in a numerical representation of the full report text and output the entire vector of predicted abnormalities, require intensive manual labeling of all abnormalities of interest for all reports in the training set, which limits the size of the training data, limits the number of abnormalities that can be predicted, and is a suboptimal choice for rare labels that will have insufficient training examples.

Instead of performing abnormality-specific whole-report classification, the first step of SARLE-Hybrid performs binary sentence classification, distinguishing only between "normal" and "abnormal" sentences rather than particular abnormalities. We define a sentence as "normal" if it describes normal findings, *e.g.*, "the lungs are clear," or the lack of abnormal findings, *e.g.*, "no masses." We define a sentence as "abnormal" if it describes the presence of abnormal findings, *e.g.*, "pneumonia in the right lung"; missing organs, *e.g.*, "thyroid is absent"; or presence of devices, lines, or tubes. We train a Fasttext model (Bojanowski et al., 2017; Joulin et al., 2017; Mikolov et al., 2013) on manually labeled sentences from the 669 training reports. This system allows sentences to be subsequently classified as indicating one or multiple abnormalities.

After the sentence classification step, a rule-based term search using medical vocabulary for each abnormality is applied to the "abnormal" sentences to determine exactly which abnormal findings are present. We designed the term search to be easily modifiable in the code for customization to other abnormalities of interest. Examples of the term search are shown in Table 2. A full description of the entire term search for all 83 abnormalities is provided in Appendix B.

The next variant, SARLE-Rules, is purely rule-based. It is identical to SARLE-Hybrid except that instead of a machine learning classifier in the first step, a rule-based system is used to identify phrases that are medically "normal" *vs.* "abnormal." The advantages of SARLE-Rules are full interpretability and better handling of the minority of sentences that include both a normal and an abnormal statement (*e.g.,* "the heart is enlarged without pericardial effusion"). The disadvantage is the extra work required to craft the rules. The rule-based phrase classifier differs from prior work in that it incorporates negation detection as well as "normality detection" based around words like "patent" (*e.g.,* "the vessels are patent"). Negation scopes are defined directly on the sentence text, include a direction (forward/backward), and can be limited by other words (*e.g.*, "and", "with") or the beginning/end of a sentence. The entirety of our "abnormality detection" including all negation detection requires fewer than 300 lines of Python code and does not have any dependencies on pretrained models (code is available at https://github.com/rachellea).

We report F-score, precision, recall, and accuracy of SARLE-Hybrid and SARLE-Rules on a held-out test set of 427 reports that were not used for classifier training or rule development. For the test reports we manually recorded abnormality-specific ground truth for 9 abnormalities commonly studied in the chest medical imaging literature: nodule, mass, opacity, consolidation, atelectasis, pleural effusion, pneumothorax, pericardial effusion, and cardiomegaly, for a total of 3,843 manual labels. The ground truth was obtained by a single observer (R.L.D., a 6th-year MD/PhD candidate). To ensure high label quality, a second observer (G.D.R, a fellowship-trained cardiothoracic radiologist) produced a random subset of 918 labels independently, resulting in 99% agreement.

We also report the F-score for the publicly available CheXpert labeler (Irvin et al., 2019) for the six CheXpert label categories that align with our label categories. We consider the CheXpert labeler's "uncertain" outputs (e.g. "possible atelectasis") as "positive" to match the way that our conservative ground truth was created.

Note that SARLE produces 83 abnormality labels per report, but due to the expense of obtaining abnormality-level ground truth, we only explicitly evaluate SARLE's performance on this subset of 9 labels. We later demonstrate the value of the additional 74 labels by showing that they improve the performance of the downstream task of multilabel CT volume classification.

**Table 2. Examples of the term search used in our radiology label extraction framework, from simple (*e.g.*, mass) to complex (*e.g.*, cardiomegaly).** The presence of any word in the "Any" column will result in considering the associated abnormality positive. The presence of any word in the "Term 1" column along with any word in the "Term 2" column will result in considering the associated abnormliaty positive. "Example Matches" shows example words and phrases that will result in a positive label for that abnormlaity based on the term search. Appendix B includes the full term search.

| Abnormality | Any | Term 1 | Term 2 | Example Matches |
|---|---|---|---|---|
| 'mass' | 'mass' | | | mass, masses |
| 'nodule' | 'nodul' | | | nodule, nodular, nodularity |
| 'opacity' | 'opaci' | | | opacity, opacities, opacification |
| 'pericardial effusion' | 'pericardial effusion','pericardial fluid' | 'pericardi' | 'fluid','effusion' | pericardial effusion present<br>effusion in the pericardial sac<br>fluid also seen in the pericardial space |
| 'cardiomegaly' | 'cardiomegaly' | 'large', 'increase', 'prominent',' dilat' | 'cardiac', 'heart', 'ventric', 'atria', 'atrium' | ventricular enlargement<br>the heart is enlarged<br>atrial dilation<br>increased heart size |

## 2.3 Development and evaluation of a whole CT volume multi-organ, multi-disease classifier

With the dataset and labels prepared, we train and evaluate a deep CNN to predict all abnormalities present in a CT volume. Following prior work on large-scale radiology datasets, we consider the automatically extracted labels ground truth (Irvin et al., 2019; Johnson et al., 2019).

Many architectures have been developed for image classification, including AlexNet (Krizhevsky et al., 2012), VGG (Simonyan and Zisserman, 2014), GoogLeNet (Szegedy et al., 2015), and ResNet (He et al., 2015). In medical imaging analysis, it is common to first pre-train one of these architectures on a large public data set of natural images (*e.g.*, ImageNet (Deng et al., 2009)) and then refine the weights on the medical images specifically, a process called "transfer learning" (Raghu et al., 2019). ResNets are a particularly popular architecture for transfer learning. ResNets include "residual connections" (also called "skip connections") which have been shown to smooth out the loss landscape and thereby facilitate training of deep networks (Li et al., 2018). Transfer learning has been shown to accelerate model convergence for medical imaging tasks (Raghu et al., 2019) including CT classification (Gao et al., 2018). However, the ResNet architecture is designed for two-dimensional images, and is thus not directly applicable to three-dimensional CT volumes.

Our proposed network architecture, refered to as CT-Net, is shown in Figure 2. First we apply a ResNet-18 feature extractor to each stack of three adjacent grayscale axial slices, which have the same shape as RGB three-channel images and can therefore serve as ResNet input. The ResNet feature extractor is pretrained on ImageNet and its weights are refined on the CT classification task. In most applications of ResNets to medical images, the classification step occurs immediately after extracting the features, using a fully-connected layer. However, because the size of a whole CT volume is so large, a fully-connected layer applied directly to the ResNet output would require 1,116,114,944 parameters. Therefore, we reduce the size of the representation by orders of magnitude and aggregate features across the whole craniocaudal extent of the data by performing 3D convolutions. Once the representation is a reasonable size, we perform the final classification using fully connected layers. To provide more insights into the model we report results on two alternative architectures and perform an ablation study.

To the best of our knowledge this is the first multilabel classification model that uses an entire CT volume as input. One prior study included a whole CT volume as a model input (Ardila et al., 2019), but the output was a lung cancer risk probability rather than predictions for multiple abnormalities. Most prior approaches have focused on 2D sections (Gao et al., 2016; Lee et al., 2019; Tang et al., 2019; Walsh et al., 2018) or patches (Anthimopoulos et al., 2016; Bermejo-Peláez et al., 2020; Christodoulidis et al., 2017;

Gao et al., 2018; Kuo et al., 2019; Li et al., 2019; Wang et al., 2019), which either requires intensive manual labeling of CT subcomponents (infeasible for a dataset of 36,316 volumes), or accepting that labels will be extremely noisy (*e.g.,* assigning the whole-volume label of "nodule" to all small patches in a CT is guaranteed to be wrong for most of the patches.)

The network is trained with a multilabel cross-entropy (CE) loss:

$$\text{CE}(y, \hat{y}) = -\frac{1}{C}\sum_{i=1}^{C}[y_i \log \hat{y}_i + (1 - y_i)\log(1 - \hat{y}_i)]$$

where $C$ is the number of abnormality labels and $y_i$ is a ground truth label for abnormality $i$. The predicted probability $\hat{y}_i$ is calculated using the logistic function (*a.k.a.* sigmoid function): $\hat{y}_i = \sigma(s_i) = \frac{1}{1+e^{-s_i}}$ for score $s_i$, where score is the raw output of the last layer. This loss function enables the model to predict multiple abnormalities per scan simultaneously.

The CT volumes vary in shape between patients. To standardize the shape we pad or center crop all CTs to shape [402, 420, 420]. We clip pixel values to [-1000, 200] Hounsfield units, normalize to the range [-1,1], and center on the ImageNet mean. The model trained for 15 days with a batch size of 2 on two NVIDIA Titan XP GPUs with 11.9 GiB of memory each (a single CT scan requires all of the memory for a single GPU, as one CT is over 1,000 times larger than a typical 256 x 256 ImageNet example). We use a stochastic gradient descent optimizer with learning rate $10^{-3}$, momentum 0.99, and weight decay $10^{-7}$. Early stopping is performed on the validation loss with patience of 15 epochs. Data augmentation is performed through random jitters to the center crop, random flips, and random rotations of an input volume. The model is implemented in PyTorch (Paszke et al., 2019). All model code is publicly available on GitHub (https://github.com/rachellea).

We train two models using the same CNN architecture (up to the last fully-connected layer): (1) a multilabel CNN trained on all 83 labels simultaneously (CT-Net-83), and (2) a multilabel CNN trained on only the 9 labels for which report-level ground truth was obtained (CT-Net-9). The intent is to demonstrate the utility of extracting multiple labels by illustrating the change in performance when the number of labels is increased from 9 to 83. We only use the test set once, for the CT-Net-83 and CT-Net-9 simultaneously, after finishing all model development.

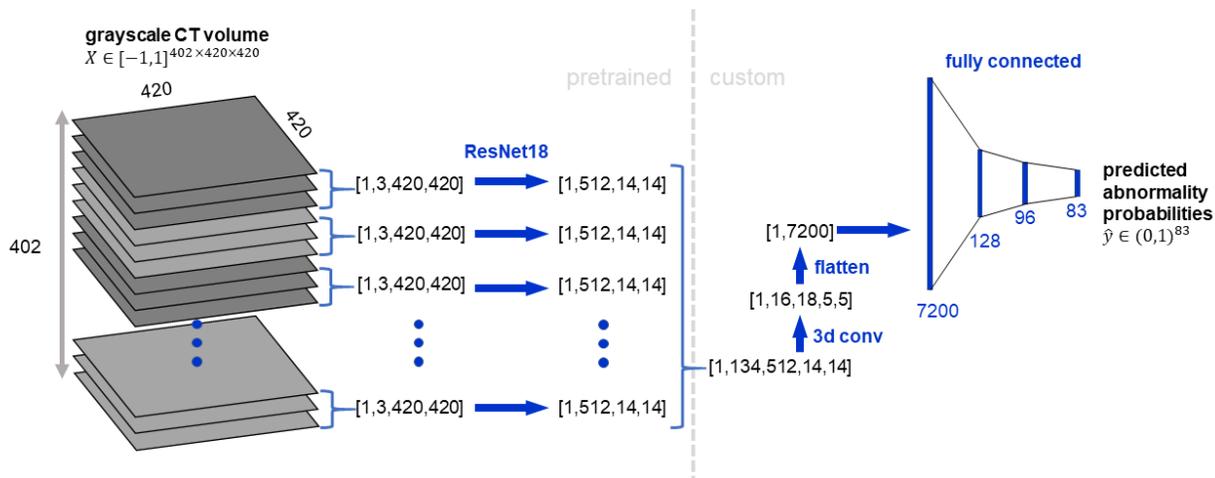

**Figure 2 CT-Net volume classification architecture.** The CT volume is treated as a stack of three-channel images to enable use of a ResNet-18 (He et al., 2015) feature extractor pretrained on ImageNet (Deng et al., 2009). The ResNet-18 features for the stack of 134 three-channel images are concatenated and processed with several 3D convolutional layers to aggregate features across the craniocaudal extent of the scan and reduce the size of the representation. Then the representation is flattened and passed through three fully connected layers to produce predicted probabilities for the 83 abnormalities of interest.

### 2.4 Architecture Comparison and Ablation Study

We compare the CT-Net architecture to two alternative architectures, BodyConv and 3DConv. BodyConv is a model similar to CT-Net-83, except instead of combining features across the whole craniocaudal extent of the scan at the beginning of the 3D convolution step, BodyConv only combines features across the craniocaudal extent at the fully connected layers. 3DConv is a model that uses only 3D convolutions and fully connected layers.

We perform an ablation study with models termed CT-Net-83 (Rand) and CT-Net-83 (Pool). The proposed CT-Net-83 architecture has three main components: the ResNet feature extractor, the 3D convolutions, and the final fully connected layers. CT-Net-83 (Rand) ablates the ResNet pretraining on ImageNet, by initializing the ResNet with random weights; this experiment was motivated by prior work (Raghu et al., 2019) suggesting that pretraining on ImageNet may not always be necessary for maximal performance on medical imaging tasks. CT-Net-83 (Pool) ablates the 3D convolution stage by replacing the 3D convolution layers with 3D max pooling layers of identical kernel size and stride. Note that we could not simply remove the 3D convolution stage entirely (*i.e.* proceed from the ResNet directly to a fully connected layer) due to the large size of the representation after the ResNet.

Performance in these experiments was obtained using a random subset of 2,000 training and 1,000 validation set scans. The full RAD-ChestCT data set of 36,316 scans was not used due to prohibitively long training/evaluation times. All these models are trained on 83 labels simultaneously.

### 2.5 Performance and Statistical Analysis

We report the area under the receiver operating characteristic (AUROC) and average precision for the CT-Net models. AUROC summarizes the sensitivity and specificity across different decision thresholds and ranges from 0.5 (random classifier) to 1.0 (perfect classifier.) Average precision is also known as the area under the precision-recall curve. While AUROC starts at a baseline of 0.5, average precision starts at a baseline equal to the frequency of positives for the particular abnormality being considered (Saito and Rehmsmeier, 2015). Therefore, an average precision of 0.4 would be high for a rare abnormality (*e.g.,* frequency 0.02) and low for a common abnormality (*e.g.,* frequency 0.8). Due to this frequency dependence we report the frequency for all abnormalities in our results.

We statistically compare the AUROCs of CT-Net-83 and CT-Net-9 using the DeLong test (DeLong et al., 1988), and obtain 95% AUROC confidence intervals using the DeLong method implemented in the pROC package in R version 3.6.2. The p-values for the DeLong test are corrected for multiple testing using the Benjamini and Hochberg method to control the false discovery rate (Benjamini and Hochberg, 1995).

## 3 Results

### 3.1 Automatic Label Extraction from Free-Text Reports

The performance of SARLE for automatic extraction of nine labels is shown in Table 3. The SARLE-Hybrid approach achieves an average F-score of 0.930 while the SARLE-Rules approach achieves an average F-score of 0.976, indicating that the automatically extracted labels are of high quality using both approaches. For the common labels, the Hybrid and Rules approaches perform equally well, *e.g.,* atelectasis where both SARLE-Hybrid and SARLE-Rules achieve an F-score of 1.0. For the rarer findings – pericardial effusion, cardiomegaly, and pneumothorax – the SARLE-Rules approach outperforms SARLE-Hybrid. Because the SARLE-Rules approach achieved higher average performance and had better rare-abnormality performance, the labels produced by SARLE-Rules were used to train and evaluate the multilabel CNN.

SARLE-Rules outperforms CheXpert on our test set of 427 reports, for the six labels that overlap between the two approaches. SARLE-Rules has a better F-score on opacity (SARLE 0.998 vs. CheXpert 0.888), consolidation (SARLE 0.975 vs. CheXpert 0.969), cardiomegaly (SARLE 0.986 vs. CheXpert 0.449), and pneumothorax (SARLE 0.941 vs. CheXpert 0.727). Both models obtain perfect performance on atelectasis. CheXpert has a better F-score than SARLE-Rules on one label, pleural effusion (SARLE 0.977 vs. CheXpert 0.983).

The most common abnormalities in the 25,355 volume training CTs are nodule (19,567 examples which is 77% positive), calcification (17,228; 68%), opacity (13,833; 55%), coronary artery disease (12,585; 50%), postsurgical (10,900; 43%), and groundglass (8,401; 33%). Note that the "nodule" category refers to any nodule, including micronodules <3 mm in size; nodules greater than 1 cm are much less frequent, at 12%. The rarest abnormalities are all at frequency 1% or less, with the following counts: hardware (321), distention (306), bronchitis (175), hemothorax (137), heart failure (50), and congestion (37). Although these abnormalities are rare, the count of positive examples for many of these abnormalities still exceeds the size of many previously reported CT data sets which are on the order of 100 – 200 CT scans total (Anthimopoulos et al., 2016; Bermejo-Peláez et al., 2020; Christodoulidis et al., 2017; Gao et al., 2018; Li et al., 2019; Wang et al., 2019).

The median number of abnormality labels for a single scan in the volume training set is 10, with an interquartile range of 6. The full histogram of abnormalities per scan is available in Appendix C. Only 139 training set scans were negative for all 83 abnormalities (*i.e.,* "normal"), which is less than 0.6% of all scans.

**Table 3: SARLE performance for the 427 chest CT test reports** across the 9 labels with manually obtained ground truth. "# Pos" is the number of positive examples for that label in the report test set. F = equally weighted harmonic mean of precision and recall, P = Precision, R = Recall, Acc = Accuracy.

| Label | # Pos | SARLE-Hybrid | | | | SARLE-Rules | | | |
|---|---|---|---|---|---|---|---|---|---|
| | | F-score | P | R | Acc | F-score | P | R | Acc |
| nodule | 341 | 0.996 | 0.991 | 1 | 0.993 | 0.996 | 0.994 | 0.997 | 0.993 |
| opacity | 213 | 0.995 | 0.995 | 0.995 | 0.995 | 0.998 | 1 | 0.995 | 0.998 |
| atelectasis | 108 | 1 | 1 | 1 | 1 | 1 | 1 | 1 | 1 |
| pleural effusion | 88 | 0.978 | 0.967 | 0.989 | 0.991 | 0.977 | 0.988 | 0.966 | 0.991 |
| consolidation | 78 | 0.969 | 0.951 | 0.987 | 0.988 | 0.975 | 0.963 | 0.987 | 0.991 |
| mass | 55 | 0.915 | 0.857 | 0.982 | 0.977 | 0.956 | 0.931 | 0.982 | 0.988 |
| pericardial effusion | 44 | 0.755 | 0.685 | 0.841 | 0.944 | 0.956 | 0.935 | 0.977 | 0.991 |
| cardiomegaly | 34 | 0.919 | 0.850 | 1 | 0.986 | 0.986 | 0.971 | 1 | 0.998 |
| pneumothorax | 8 | 0.842 | 0.727 | 1 | 0.993 | 0.941 | 0.889 | 1 | 0.998 |

## 3.2 Multilabel CNN Predicts Abnormalities from CT Volumes

Table 4 compares the test set performance of the CT-Net-83 and CT-Net-9 models. The CT-Net-83 model outperforms the CT-Net-9 model on every abnormality, indicating the value of training on the additional 74 labels. This is an example of the benefit of transfer learning.

Table 5 shows the abnormality labels for which the CT-Net-83 model achieved the highest and lowest performance. Several of the highest-performing labels are abnormalities related to surgeries that affect a large area of the chest, including lung resection, sternotomy, CABG (coronary artery bypass graft), transplant, and "postsurgical" which encompasses a variety of descriptors of recent surgery. Other high-performing labels are human-made objects, including pacemaker or defibrillator, tracheal tube, catheter or port, heart valve replacement, chest tube, and GI tube. Finally, there are numerous common biological abnormalities that the model is able to identify with high AUROC, including pleural effusion, emphysema, pulmonary edema, fibrosis, interstitial lung disease, pneumothorax, and coronary artery disease.

The model performs poorly on several labels, including cyst, density, and scattered nodules (Table 5). On further analysis we discovered that cysts most commonly appear in the kidneys, which are likely to be cropped out in the preprocessing due to appearing at the edge of the volume. "Density" is used in variable ways in radiology reports and may not correspond to a clear visual pattern. The "scattered nodules/nodes" category includes scattered micronodules which may affect only one or two pixels each and by definition are distributed over a wide area, which may be a difficult characteristic for the model to capture.

Overall, CT-Net-83 achieves an AUROC >0.9 on 18 abnormalities, 0.9 – 0.8 AUROC on 17 abnormalities, 0.8 – 0.7 AUROC on 24 abnormalities, 0.7 – 0.6 AUROC on 18 abnormalities, and <0.6 AUROC on 6 abnormalities. Performance of CT-Net-83 on all 83 abnormalities is provided in Appendix C.

In Figure 3 we present a boxplot summary of the AUROC across all 83 abnormalities for our proposed CT-Net-83 architecture and two alternative architectures on a validation set of 1,000 volumes for models trained on 2,000 volumes. CT-Net-83 in this figure shows the performance of the CT-Net-83 model on this smaller subset of data; the performance is lower than that shown in Tables 4 and 5, due to training on 12× less data. CT-Net-83 outperforms the two alternative architectures, BodyConv (which uses different 3D convolutions) and 3DConv (which uses all 3D convolutions instead of a pre-trained ResNet feature extractor).

Figure 3 also demonstrates that the proposed CT-Net-83 outperforms the two ablated variants, (Rand) in which the ResNet feature extractor is randomly initialized instead of pretrained, and (Pool) in which the 3D convolutions have been replaced by max pooling operations. These results illustrate that for the task of multiple abnormality prediction in CT scans, transfer learning by pretraining the feature extractor on ImageNet does lead to better abnormality identification in spite of the differences between CT slices and natural images. Furthermore, the relationships learned in the 3D convolution stage are critical for the model's performance. Without the 3D convolutions, the model does not converge.

**Table 4. CT volume test set AUROC for models trained on 9 vs. 83 labels.** The area under the receiver operating characteristic (AUROC) is shown for CT-Net-9 (trained only on the 9 labels shown) and CT-Net-83 (trained on the 9 labels shown plus 74 additional labels) for the test set of 7,209 examples. CT-Net-83 outperforms CT-Net-9 on all abnormalities, emphasizing the value of the additional 74 labels. Note that we also experimented with separate binary classifiers for each of the 9 labels independently, but these models did not converge (AUROC ~0.5). Positive Count and Positive Percent are for positive examples of the abnormality in the test set.

| Abnormality | Positive Count | Positive Percent | CT-Net-9 AUROC | 95% CI | CT-Net-83 AUROC | 95% CI | DeLong p-value |
|---|---|---|---|---|---|---|---|
| nodule | 5,617 | 77.9 | 0.682 | 0.667-0.698 | **0.718** | 0.703-0.732 | $3.346 \times 10^{-7}$ |
| opacity | 3,877 | 53.8 | 0.617 | 0.605-0.630 | **0.740** | 0.728-0.751 | $<4.950 \times 10^{-16}$ |
| atelectasis | 2,037 | 28.3 | 0.683 | 0.668-0.697 | **0.765** | 0.753-0.777 | $<4.950 \times 10^{-16}$ |
| pleural effusion | 1,404 | 19.5 | 0.945 | 0.937-0.952 | **0.951** | 0.945-0.958 | $1.882 \times 10^{-2}$ |
| consolidation | 1,086 | 15.1 | 0.719 | 0.703-0.736 | **0.816** | 0.804-0.829 | $<4.950 \times 10^{-16}$ |
| mass | 863 | 12.0 | 0.624 | 0.604-0.644 | **0.773** | 0.755-0.791 | $<4.950 \times 10^{-16}$ |
| pericardial eff. | 1,078 | 15.0 | 0.659 | 0.640-0.677 | **0.697** | 0.679-0.714 | $8.315 \times 10^{-8}$ |
| cardiomegaly | 649 | 9.0 | 0.791 | 0.774-0.807 | **0.851** | 0.836-0.867 | $7.000 \times 10^{-13}$ |
| pneumothorax | 205 | 2.8 | 0.816 | 0.785-0.847 | **0.904** | 0.882-0.926 | $8.810 \times 10^{-11}$ |

**Table 5. CT-Net-83 test set AUROC and average precision for abnormalities with the highest and lowest AUROCs.** Note that the baseline for average precision is equal to the frequency of the abnormality being considered; this frequency is provided in the Test Set Percent column. Thus, an average precision of 0.463 for honeycombing is high, given honeycombing's baseline of only 0.027.

| Abnormality | AUROC | Average Precision | Test Set Percent | Test Set Count |
|---|---|---|---|---|
| pacemaker or defib | 0.975 | 0.699 | 0.039 | 279 |
| honeycombing | 0.972 | 0.463 | 0.027 | 193 |
| tracheal tube | 0.971 | 0.597 | 0.017 | 121 |
| lung resection | 0.967 | 0.876 | 0.194 | 1,398 |
| sternotomy | 0.965 | 0.598 | 0.071 | 514 |
| CABG | 0.965 | 0.527 | 0.040 | 288 |
| transplant | 0.963 | 0.751 | 0.057 | 414 |
| catheter or port | 0.954 | 0.716 | 0.105 | 755 |
| heart failure | 0.952 | 0.040 | 0.002 | 18 |
| pleural effusion | 0.951 | 0.869 | 0.195 | 1,404 |
| heart valve replacement | 0.949 | 0.219 | 0.018 | 133 |
| chest tube | 0.944 | 0.387 | 0.020 | 146 |
| GI tube | 0.940 | 0.492 | 0.022 | 162 |
| emphysema | 0.929 | 0.843 | 0.243 | 1,754 |
| pulmonary edema | 0.921 | 0.524 | 0.052 | 373 |
| fibrosis | 0.910 | 0.662 | 0.112 | 811 |
| interstitial lung disease | 0.906 | 0.764 | 0.153 | 1,102 |
| pneumothorax | 0.904 | 0.355 | 0.028 | 205 |
| postsurgical | 0.896 | 0.853 | 0.428 | 3,089 |
| hemothorax | 0.890 | 0.038 | 0.004 | 28 |
| coronary artery disease | 0.873 | 0.830 | 0.494 | 3,563 |
| cyst | 0.594 | 0.184 | 0.143 | 1,032 |
| granuloma | 0.588 | 0.116 | 0.083 | 595 |
| hardware | 0.577 | 0.020 | 0.017 | 120 |
| density | 0.560 | 0.115 | 0.090 | 647 |
| scattered nodules/nodes | 0.559 | 0.249 | 0.210 | 1,512 |
| infiltrate | 0.526 | 0.016 | 0.015 | 107 |

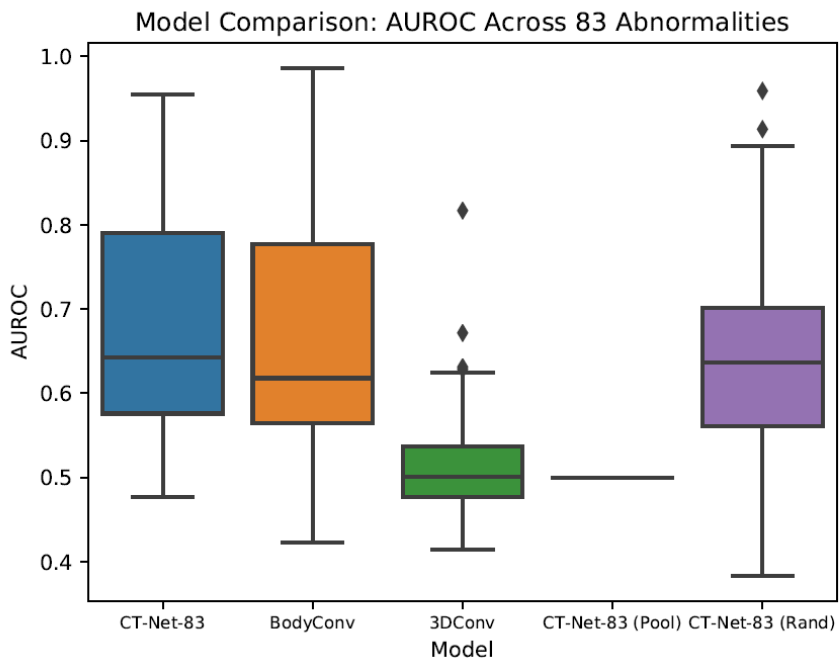

**Figure 3. Architecture Comparison and Ablation Study on Training/Validation Data Subset.**
The AUROCs for each abnormality in this experiment were calculated on a random sample of 1,000 validation set scans, for models trained on a random subset of 2,000 training scans. CT-Net-83 is the proposed model. BodyConv and 3DConv are alternative architectures. CT-Net-83 (Pool) and CT-Net-83 (Rand) are ablated version of the CT-Net-83 model.

## *4 Discussion*

The three main contributions of this work are the preparation of the Report-Annotated Duke Chest CT data set (RAD-ChestCT) of 36,316 unenhanced chest CT volumes, the SARLE framework for automatic extraction of 83 labels from free-text radiology reports, and a deep CNN model for multiple abnormality prediction from chest CT volumes.

The RAD-ChestCT data set is the largest reported data set of multiply annotated chest CT volumes, with 36,316 whole volumes from 19,993 unique patients. We plan to make the CT volumes publicly available, pending deidentification and approval. Fewer than 6% of studies on deep learning in radiology use more than 10,000 cases (Soffer et al., 2019). Several previous studies on interstitial lung disease use between 120 and 1,157 chest CTs (Anthimopoulos et al., 2016; Bermejo-Peláez et al., 2020; Christodoulidis et al., 2017; Gao et al., 2018, 2016; Walsh et al., 2018; Wang et al., 2019). Two recent studies on acute intracranial hemorrhage used 1,300 (Lee et al., 2019) and 4,596 (Kuo et al., 2019) head CTs. The public LIDC-IRDI data set (Armato et al., 2011) of 1,018 chest CT volumes and the public DeepLesion data set (Yan et al., 2018) of 10,825 partial CT volumes are centered on focal lesions (*e.g.,* nodules). Ardila et al. (Ardila et al., 2019) develop a lung cancer screening model on a National Lung Screening Trial (NLST) (Gatsonis et al., 2011) data set of 42,290 CT scans with cancer-related annotations, from 14,851 patients. This represents a greater total number of scans, but a smaller number of unique patients and unique annotations than RAD-ChestCT.

To the best of our knowledge RAD-ChestCT is the only chest CT data set with such a diverse range of abnormality annotations including both focal (nodule, mass, *etc.).* and diffuse (fibrosis, ILD, atelectasis, edema, pneumonia, *etc.)* abnormalities. In its present form it can be used for multilabel classification, weakly supervised abnormality localization, or exploratory research using unsupervised methods like clustering. One potential future direction would be to extend the RAD-ChestCT data set to include bounding box annotations to facilitate supervised abnormality localization. To accelerate other large-scale machine learning projects on CT data, we provide a detailed tutorial in Appendix A on how to transform raw CT DICOMs into 3D numpy arrays for analysis, and we have made our entire end-to-end Python CT preprocessing pipeline publicly available.

The SARLE framework for automatic label extraction from radiology reports is designed to be simple, scale to a large number of abnormality labels, and achieve high performance. It is the first approach to automatically extract numerous abnormality labels from chest CT reports. The general principle of first distinguishing between medically "normal" and "abnormal" phrases, and then performing an abnormality-specific vocabulary lookup, is applicable to any radiology modality. In Appendix B we present a detailed discussion of SARLE in the context of related work. Other studies in label extraction report F-scores between 0.52 – 1.0 for the extraction of anywhere between 3 and 55 abnormalities (Banerjee et al., 2017; Chen et al., 2018; Demner-Fushman et al., 2016; Irvin et al., 2019; Peng et al., 2018; Pham et al., 2014; Zech et al., 2018). Relative to these other approaches, SARLE achieves high F-score (average 0.976 across 9 abnormalities) and makes predictions on a large number of labels (83). We found that SARLE-Rules outperforms the CheXpert labeler on our test set of chest CT reports. This may be in part because the CheXpert labeler's rules and vocabulary were originally designed for chest x-ray reports rather than chest CT reports. We also found that SARLE-Rules outperforms SARLE-Hybrid, and hypothesize that this may be because the Rules approach is phrase-based rather than whole-sentence based, which allows better handling of the small minority of sentences that contain both normal and abnormal medical findings. In the future, the SARLE-Hybrid approach could be extended by replacing the sentence classifier with a phrase classifier.

Our work was inspired by the ChestX-Ray8 study (Wang et al., 2017) and the CheXpert study (Irvin et al., 2019), which share the most similar overall experimental design. These studies as well as our present work involve preparation of a large database of radiology images, development and application of an automated label extraction approach to obtain structured disease annotations from radiology reports, and training and evaluation of a deep learning model on the radiology images using the automatically extracted labels as ground truth.

However, there are a few key differences between our work and the ChestX-Ray8/CheXpert studies. First, ChestX-Ray8/CheXpert are focused on projectional radiographs, which are two-dimensional images, while our work focuses on CT scans, which are three-dimensional and require different preprocessing steps and modeling considerations due to their volumetric nature. Our label extraction approaches also differ: the CheXpert work applies rules defined on a sentence graph to extract 14 abnormalities with F-scores ranging between 0.72 – 1.00 while our work uses rules defined directly on the sentence text to produce 83 labels per report, with F-scores for 9 abnormalities ranging between 0.941–1.00.

The challenges in applying deep learning models to radiographs and CTs are different. A single CT is about 70x larger than a radiograph, which presents hardware and memory challenges and causes the abnormalities to be more spatially dispersed within the training example. Because radiographs are 2D projections of a 3D volume, radiographs are more ambiguous (de Hoop et al., 2010; Gibbs et al., 2007; Howarth and Tack, 2015; Self et al., 2013), which alters what abnormalities can be visualized and changes the implications of certain words in the reports. For example, the majority of nodules smaller than 1 centimeter are not visible on chest radiographs (MacMahon et al., 2017) whereas CT can detect nodules as small as 1 – 2 mm in diameter (Sánchez et al., 2018). That means our "nodule" category for CT includes nodules that from a volumetric perspective are up to 1,000× smaller than those visible on chest radiographs, and these tiny CT nodules are distributed across 1,000× more pixels. In spite of this, we obtain better performance on nodule and mass identification: nodule 0.718 ours *vs.* 0.716 ChestX-Ray8 (Wang et al., 2017), and mass 0.773 *vs.* 0.564 ChestX-Ray8.

Although the studies are on different kinds of data using different automatic labelers, for the purposes of placing our work in context, we note that our CT volume classifier's performance is generally in the same range as that of prior chest x-ray work: pneumothorax 0.904 ours *vs.* 0.789 ChestX-Ray8 (Wang et al., 2017), pneumonia 0.816 ours *vs.* 0.633 ChestX-Ray8, infiltration 0.526 ours *vs.* 0.612 ChestX-Ray8. The CheXpert study (Irvin et al., 2019) reports performance on additional abnormalities: atelectasis 0.765 ours *vs.* 0.858 CheXpert, cardiomegaly 0.851 ours *vs.* 0.832, consolidation 0.816 ours *vs.* 0.899, edema 0.921 ours *vs.* 0.941, and pleural effusion 0.951 ours *vs.* 0.934.

Most prior research in CT scan classification has focused on a single category of diseases and relies on a fundamentally different modeling approach that requires manual slice-level, patch-level, or pixel-level labels. Several studies have focused on subtypes of interstitial lung disease (ILD) (Anthimopoulos et al., 2016; Bermejo-Peláez et al., 2020; Christodoulidis et al., 2017; Gao et al., 2018, 2016; Walsh et al., 2018) but all of these studies require manual labeling of regions of interest or manual pixel-level labeling, and all the models are trained on small patches (*e.g.*, $32 \times 32$ pixels) or slices extracted from the CT scan. The advantage of patch or slice classification is that it provides some inherent localization. The disadvantage is that it limits the total number of CTs in the data set (all these studies use <1,200 CTs) and it limits the total number of abnormalities that can be considered (all consider <9 classes) due to the immense manual work required to obtain pixel-level, patch-level, or slice-level annotations.

A related study (Tang et al., 2019) performs binary classification of weakly-labeled slices for each of nodule, atelectasis, edema, and pneumonia separately versus normal scans, using an approximately 1:1 ratio between abnormal and normal scans. This is a different task from our multilabel setup, in which we use unfiltered hospital data where fewer than 1% of scans are normal, and we train one model on all abnormalities simultaneously. Two implications of our approach are noteworthy. Firstly, the application of multiple single label models does not provide a basis for accommodating interactions between co-existent diseases or imaging findings. Our multilabel approach provides a holistic image assessment paradigm that may better generalize across individual patients by accommodating interactions across the clinical reality of multiple co-existent diseases and imaging findings. Secondly, the very low prevalence of true "normality" in an unfiltered sample of patients receiving unenhanced chest CT scans challenges the utility of interpretation workflows that are based upon patient-level prioritization of abnormal scans over those that are normal.

A substantial body of literature focuses on lung nodules and is the subject of multiple review articles (Pehrson et al., 2019; Shaukat et al., 2019; Zhang et al., 2018). For understandable reasons these methods typically rely on manually acquired nodule bounding boxes and can achieve AUROCs in the upper 0.90s. This is higher than our AUROC of 0.718 for nodules, but our model was trained using only whole-volume labels without any bounding box annotations. In the future it may be possible to improve our model's performance on focal findings through combined training on RAD-ChestCT and DeepLesion or LIDC, through addition of bounding box annotations to RAD-ChestCT, or through active learning to improve a weakly-supervised lesion detector. Multi-scale architectures may facilitate improvement on both focal and diffuse findings simultaneously.

A critical contribution of our work is the demonstration that leveraging numerous automatically extracted abnormality labels enables learning from unfiltered hospital-scale CT data. A binary classifier trained on unfiltered hospital CT data does not converge (AUROC ~0.5) likely due to contamination of the "normal" class with other abnormalities, some of which may look similar to the target class. Training a multilabel classification model on 83 labels simultaneously instead of only 9 simultaneously boosts the average AUROC by over 10%, from 0.726 to 0.802. Furthermore, the model trained on all 83 labels achieves AUROCs over 0.90 for almost twenty different abnormalities including many medically significant abnormalities that have been the subject of prior work (Anthimopoulos et al., 2016; Christe et al., 2019; Humphries et al., 2019; Irvin et al., 2019; Li et al., 2019; Tang et al., 2019; Walsh et al., 2018) such as emphysema (0.929), pleural effusion (0.951), pulmonary edema (0.921), interstitial lung disease (0.906), honeycombing (0.972), pneumothorax (0.904), and fibrosis (0.910). We further show that our proposed model, CT-Net, is able to outperform two alternative models by leveraging transfer learning and 3D convolutions that combine abnormality features across the craniocaudal extent of the scan.

We hope this work will contribute to the long-term goal of augmented medical image interpretation systems that enhance the radiologists' workflow, improve detection and monitoring of diseases, and advance patient care.

## *Declaration of Competing Interest*

The authors declare that they have no known competing financial interests or personal relationships that could have appeared to influence the work reported in this paper.

## *Acknowledgements*

We would like to thank Mark Martin, Justin Solomon, the Duke University Office of Information Technology (OIT), and the Duke Protected Analytics Computing Environment (PACE) team. We also thank the anonymous reviewers for providing insightful comments that improved the manuscript.

## *Funding Sources*

This work was supported by NIH/NIBIB R01-EB025020, developmental funds of the Duke Cancer Institute from the NIH/NCI P30-CA014236 Cancer Center Support Grant, and GM-007171 the Duke Medical Scientist Training Program Training Grant.

# Appendix A. CT Data Preparation

## A.1 CT Report Preparation

### A.1.1 Background & Downloading CT Reports

A CT report is associated with three important fields: the medical record number (MRN), the accession number, and the protocol description.

The MRN uniquely identifies the patients. The MRN is critical for defining a train/validation/test split in which no patient appears in more than one set. CT scans from the same patient must all be grouped into the same set because even if the scans occurred years apart, they are still correlated because they depict the same person.

The accession number uniquely identifies the CT scan event. It is an alphanumeric identifier (*e.g.,* "AA12345") distinct from the MRN that specifies a particular occurrence of a particular patient obtaining a CT scan. The accession number of a CT scan event is used for both the CT report and the CT volume associated with that event. The accession number is necessary for matching up reports with their corresponding CT volumes.

The protocol description is a string that describes what body part was imaged, whether contrast was used, and other pertinent information the scan acquisition. More details about CT protocols is included in section A.1.4.

Initially, 414,438 CT reports were downloaded using the Duke Enterprise Data Unified Content Explorer (DEDUCE), which is a tool that enables clinicians to obtain data from Duke's electronic health record (Horvath et al., 2011). An additional 26,384 reports were downloaded directly from the Epic electronic health record.

### A.1.2 Removing Duplicate CT Reports

The raw data set of 414,438 CT reports from DEDUCE contained some duplicates. Table A1 summarizes the process of removing these duplicates. The final number of unique reports from DEDUCE was 330,710.

**Table A1. Removing duplicates from the raw CT report data.**

| Step | New Reports Count After Step |
|---|---|
| **Raw data** | 414,438 |
| **Exact duplicates:** drop exact duplicates. | 412,947 |
| **Preliminary vs. verified**: For all reports with at least one verified report, drop all preliminary reports. | 331,842 |
| **Addended or not**: For all reports with at least one addended version, drop all un-addended versions. | 330,949 |
| **Multiple addendums**: For reports with at least one multiply addended version, drop all versions with fewer addenda. | 330,930 |
| **Human error:** Manually remove duplicate reports introduced by human error. | 330,917 |
| **Empty reports:** Remove empty reports (those with <550 characters). | 330,710 |

### A.1.3 Merging CT Report Datasets

The 330,710 DEDUCE reports were merged with the 26,384 Epic reports. 77% of the Epic reports overlapped with reports in the DEDUCE data set. The remaining 23% of the Epic reports (6,093 reports) were unique to Epic. This is due to slight differences in the databases used by the Epic electronic health record and the DEDUCE tool. After merging of DEDUCE and Epic reports and removal of rows with missing values, the final merged report data set included 336,800 unique reports.

### A.1.4 Filtering by Protocol

A CT protocol is a particular recipe for obtaining a CT scan. A CT protocol has several components.

The first component of a CT protocol is the specification of the part of the body to be imaged. Different locations include chest only, abdomen only, pelvis only, chest/abdomen/pelvis together, head, and spine. If a patient has a suspicious lung nodule on chest x-ray, a follow-up chest CT scan may be ordered to better evaluate the nodule(MacMahon et al., 2017). If a patient has symptoms of a stroke, a head CT may be ordered (Birenbaum et al., 2011). If a patient suffers a car accident, a CT of their cervical spine may be needed to determine whether their neck is fractured (Wee et al., 2008). Clinicians are careful to only order a CT scan of the part of the body relevant to the individual patient's disease, to limit the patient's dose of x-ray radiation (Costello et al., 2013). Thus, the body location narrows down the possible reasons for that particular CT scan, which means that the location specified in the CT changes which abnormalities may be present in that CT.

A second important component of a CT protocol is the description of whether contrast was used, and how. Contrast is a radiopaque liquid typically made from barium or iodine that is injected into arteries, injected into veins, delivered by enema, or swallowed by the patient in order to highlight particular structures (Lusic and Grinstaff, 2013). On a CT scan, contrast appears bright white. The type of contrast agent, the method of administration, and the timing relative to the acquisition of the scan can be specified in the CT protocol. A CT pulmonary angiogram is a type of chest CT in which contrast is injected to fill the pulmonary blood vessels; it is often used for the diagnosis of pulmonary embolism, which is a clot in the pulmonary blood vessels. The clot appears greyer than the surrounding white contrast and is thus easier for the radiologist to identify (Wittram et al., 2004). Contrast CTs of the abdomen and pelvis can be obtained to evaluate for appendicitis, diverticulitis, and pancreatitis (Rawson and Pelletier, 2013). In other cases, contrast is unnecessary, or even contraindicated (*e.g.,* if the patient has kidney disease, is pregnant, or has a past history of negative reactions to contrast agents.)

In total, there are 458 different protocols included in the Duke CT reports dataset. Because different protocols encompass different parts of the body, different medical motivations, and different appearance of the same anatomical structures (depending on whether and how contrast is used), it would be difficult to analyze all CT scans in the same machine learning model. Not even humans attempt to be experts in interpretation of images from all parts of the body. There are numerous subspecialties within the field of radiology that focus on particular organ systems, including neuroradiology, cardiovascular radiology, gastrointestinal radiology, musculoskeletal radiology, and head and neck radiology ("Diagnostic Radiology Professions," n.d.). We choose to focus on chest CTs obtained without intravenous contrast material.

The top 10 commonest protocols in the complete Duke CT reports data set are shown in Table A2. Our chosen protocol, for chest CTs without intravenous contrast, is the second most common and makes up 9.8% of the total data set. Chest CTs without intravenous contrast can be used to evaluate a huge range of medical conditions including solitary pulmonary nodules, interstitial lung disease, pleural effusions, lung cancer, infections, inflammation, and edema (Purysko et al., 2016).

To create our final selection of chest CTs obtained without intravenous contrast, we extracted all reports for which the lowercased protocol description matched either "ct chest wo contrast w 3d mips protocol" or "ct chest without contrast with 3d mips protocol." The count of reports matching either of these protocol descriptions was 36,861, which is 11% of the total report data set.

**Table A2. The top 10 most common protocols in the complete Duke CT reports data set, 2011 – 2017.** The main protocol selected for creation of the RAD-ChestCT data set is bolded: "CT chest wo contrast w 3D MIPS Protocol."

| Protocol | Count (out of 336,800 total) |
| --- | --- |
| CT abdomen pelvis with contrast | 43,904 |
| **CT chest wo contrast w 3D MIPS Protocol** | 32,883 |
| CT chest abdomen pelvis with contrast w MIPS | 30,619 |
| CT ABDOMEN PELVIS W CONT | 17,929 |
| CT abdomen pelvis without contrast | 14,897 |
| CT brain without contrast | 13,137 |
| CT chest PE protocol incl CT angiogram chest w wo contrast | 12,064 |
| CT CHEST W/ENHANCE | 11,683 |
| CT CHEST | 11,268 |
| CT MIPS | 10,764 |

### A.1.5 Cleaning CT Reports

All CT reports were prepared for analysis with standard NLP preprocessing techniques. This preprocessing included extracting both the "Findings" and "Impression" sections of the report, deleting the "Findings" and "Impression" section headers if present (to avoid thousands of sentences beginning with those words), splitting the report by sentence, lowercasing, and replacing all whitespace with a single space. We also removed punctuation except for the periods inside decimal numbers, which carry medical meaning (*e.g.* 1.2 cm mass versus 12 cm mass). Finally, we replaced all times with a "%time" token, dates with "%date", and years with "%year."

## *A.2 CT Volume Preparation*

### A.2.1 Downloading CT Volumes

All 36,861 selected chest CT accession numbers were queried using an API developed for the Duke Image Archive. The download process required 49 days and had a 99% success rate, with 36,316 volumes acquired out of 36,861 initially specified.

### A.2.2 The Data Interchange Standard for Biomedical Imaging (DICOM)

CT scans are stored using the DICOM format(Bidgood et al., 1997). An image in DICOM format is saved as a pixel array with associated metadata. The metadata includes information about the patient, including patient name, ID, birth date, and gender. The metadata also includes information about the scan itself. For the purposes of creating the RAD-ChestCT dataset, several DICOM metadata attributes were particularly important. These attributes are summarized in Table A3.

**Table A3. DICOM Attributes Important for CT Volume Preprocessing.** All entries in the Description column are excerpts from the DICOM Standard Browser by Innolitics("Image Type Attribute – DICOM Standard Browser," n.d.).

| DICOM Attribute | Description | Usage |
| --- | --- | --- |
| Image Type | Image Type (0008,0008) identifies important image identification characteristics. These characteristics are: (a) Pixel Data Characteristics: 1. is the image an ORIGINAL Image; an image whose pixel values are based on original or source data; 2. is the image a DERIVED Image; an image whose pixel values have been derived in some manner from the pixel value of one or more other images. (b) Patient Examination Characteristics: 1. is the image a PRIMARY Image; an image created as a direct result of the patient examination; 2. is the image a SECONDARY Image; an image created after the initial patient examination. | Choose the series that is ORIGINAL. |
| Image Orientation (Patient) | The direction cosines of the first row and the first column with respect to the patient. | Ensure orientation '1,0,0,0,1,0'. |
| Image Position (Patient) | The x, y, and z coordinates of the upper left-hand corner (center of the first voxel transmitted) of the image, in mm. | Ensure correct ordering of 2D slices to make 3D volume. |
| Rescale Intercept | The value b in relationship between stored values (SV) and the output units. Output units = m*SV+b If Image Type (0008,0008) Value 1 is ORIGINAL and Value 3 is not LOCALIZER, output units shall be Hounsfield Units (HU). | Convert pixel values to Hounsfield units. |
| Rescale Slope | m in the equation specified in Rescale Intercept (0028,1052). | |
| Pixel Spacing | Physical distance in the patient between the center of each pixel, specified by a numeric pair - adjacent row spacing (delimiter) adjacent column spacing in mm. | Ensure consistent pixel distance meaning. |
| Gantry/Detector Tilt | Nominal angle of tilt in degrees of the scanning gantry. | Ensure zero tilt. |

### A.2.3 Selecting Series

A single CT scanning event is indicated by one unique accession number, but it might include more than one CT volume. For example, if the first scan is of insufficient quality, a second scan may be obtained. Furthermore, once the scanning is finished, the radiologists who interpret the scans may require "reformats" which are alternative computationally acquired representations of the scan data that display the patient's body from a different view. Each separate CT scan obtained during the CT scanning event and each separate reformat created afterwards is referred to as a separate "series."

We defined a process for choosing which series to use as the definitive CT volume representing a particular accession number. We first made use of the *ImageType* DICOM attribute (Table A3) so that we only considered scans which were marked as ORIGINAL, excluding scans marked as DERIVED, SECONDARY, or REFORMATTED. Next, for accession numbers with more than one ORIGINAL series, we selected the series with the greatest number of slices.

### A.2.4 Ordering Slices

Separate DICOM files are saved for each axial slice of a CT volume. Relative to a standing person, an "axial slice" represents a horizontal plane through the body (the same plane formed by a belt around the waist). Figure A1 illustrates the axial slice plane.

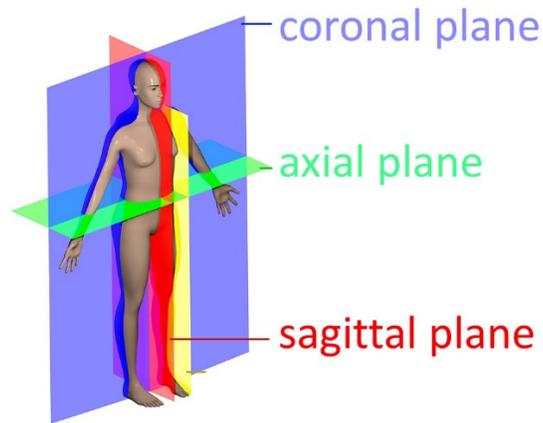

**Figure A1 Coronal, axial, and sagittal planes.** CT scans are saved using axial slices. Figure adapted from [Wikipedia original figure by David Richfield and Mikael Häggström, M.D.(Richfield and Haggstrom, n.d.)](#)

The separate axial slices must be stacked in the correct order in order to recreate the 3D volume – a process which is surprisingly nontrivial, but critical in order to obtain usable data. If slices are shuffled the volumetric representation is destroyed.

The most intuitive DICOM attribute to use for slice ordering is called *InstanceNumber*. This attribute is supposed to specify the order of the slices using integers. Unfortunately, this attribute is not reliable and is filled incorrectly by some CT scanners. Therefore, we did not use the *InstanceNumber* attribute.

The most reliable way to order slices is to make use of the *ImageOrientationPatient* and *ImagePositionPatient* attributes (Table A3). Figure A2 depicts the 12 possible orientations of a patient in a volume. *ImageOrientationPatient* specifies the patient's orientation and is needed in order to determine which direction is the "z" direction (*a.k.a.* craniocaudal direction). The "z" direction is the direction along which slices must be stacked. Typically, patients are presented in the "1,0,0,0,1,0" orientation for a chest CT, but the orientation is still important to verify as it can sometimes vary.

Given the patient orientation we can determine which value stored in the *ImagePositionPatient* attribute is the z-position. The z-position reflects a physical distance measurement of the craniocaudal location of the slice in space. By sorting slices according to their z-position, it is possible to obtain the correct slice ordering and reconstruct the full 3D volume.

### A.2.5 Rescaling Pixel Values to Hounsfield Units

Raw pixel values in DICOMs have undergone a linear transformation to enable efficient disk storage. This transformation must be reversed to obtain pixel values in Hounsfield units (HU). The DICOM standard includes the attributes *RescaleSlope* and *RescaleIntercept* which are the "$m$" and the "$b$" in the equation $y = mx + b$, needed to transform the raw pixel values $x$ stored in the DICOM pixel array into Hounsfield units $y$. We rescaled all raw pixel values to HUs using the *RescaleSlope* and *RescaleIntercept* attributes. This ensures that a particular numerical value for a pixel indicates the same radiodensity across all scans.

After converting to HU, we clipped pixel values to [-1000 HU,+1000 HU], which represent practical lower and upper limits of the HU scale, corresponding to the radiodensities of air and bone respectively (DenOtter and Schubert, 2019). Water has a radiodensity of 0 HU and most tissues are in the range -600 to +100 HU (Lamba et al., 2014). When training our model, we dynamically clip the pixel values to [-1000 HU, +200 HU] since we focus mainly on heart and lung abnormalities. We store the wider range of [-1000 HU,+1000 HU] in case future work would benefit from higher regions of the HU scale, *e.g.* future work focused on bones.

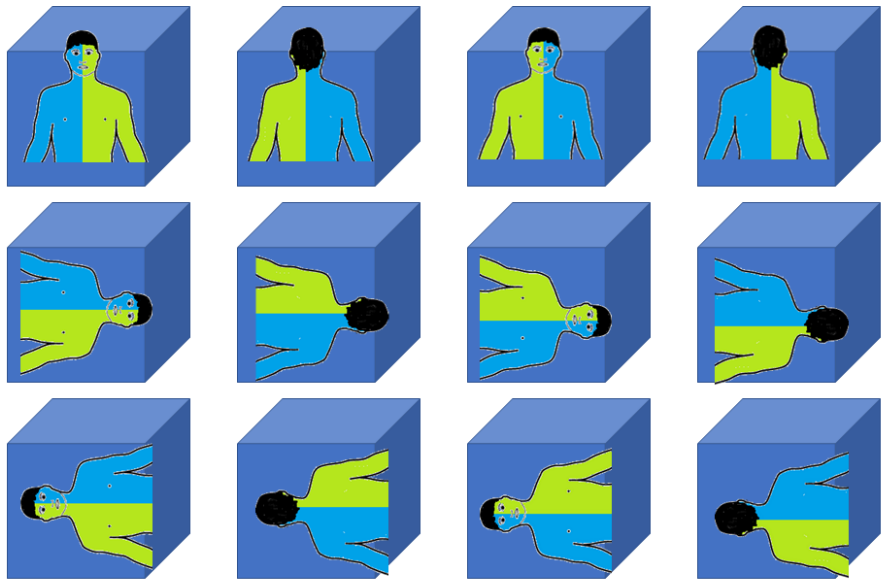

**Figure A2 Twelve possible orientations for a human in a cube.** Orientation is important because humans are asymmetrical on the inside (*e.g.* the heart is towards the left, the stomach is on the left, the liver is on the right). The human figure sketch is adapted from Wikipedia("File:Human_body_schemes.png - Wikimedia Commons," n.d.).

**A.2.6 Resampling**

In addition to ensuring that the pixel values have consistent radiodensity meaning across the whole data set by converting to Hounsfield units, we also need to ensure that each pixel represents a consistent volume throughout the whole data set. A DICOM header includes the attribute *PixelSpacing* (Table A3) which includes an *x* measurement and a *y* measurement in millimeters (*x*-spacing and *y*-spacing). A *z*-spacing (craniocaudal spacing) can be inferred by subtracting the *z*-position values of adjacent slices reported in *ImagePositionPatient*. Taken together, (*x,y,z*) spacing indicates the physical distance in millimeters that a single pixel represents.

Unfortunately, these physical distances vary depending on the scan parameters, such that the physical size of each pixel is often different for different patients (Figure A3). In our data set, 19,619 downloaded volumes had *z*-spacing of 0.625 mm (calculated from *ImagePositionPatient*) and 14,871 volumes had 0.6 mm *z*-spacing, together representing 95% of all volumes. Remaining volumes were divided between those with *z*-spacing <0.6 mm (23) and >0.625 mm (1,803). Note that in a given scan the *x*- and *y*-spacing values are typically different from the *z*-spacing value. For example, the *x*- and *y*-spacing values may be 0.732 mm or 0.793 mm as shown in Figure A3.

To achieve consistent physical distance meaning of all pixels in the data set, we resampled all CTs to 0.8 × 0.8 × 0.8 mm using the Python package Simple ITK (Lowekamp et al., 2013). This step also has the advantage of mildly decreasing the size of each volume, which reduces disk storage requirements and memory requirements of machine learning models trained on this data set.

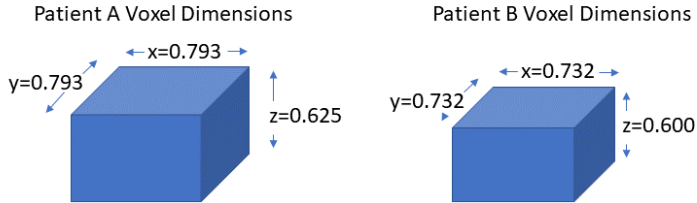

**Figure A3 Conceptual example of different ($x,y,z$) pixel spacing between two patients in the data set.** The values shown are real examples of different $x$, $y$, and $z$ values from the data set.

### A.2.7 Resampling with Irregular Z-Spacing

For a single CT scan, a $z$-spacing value is calculated between each pair of adjacent axial slices. Thus, for a typical CT scan with hundreds of slices, there are hundreds of opportunities for irregularities in $z$-spacing. Approximately 4% of total scans have at least one instance of irregular $z$-spacing between any of their slices.

It is important to handle these scans carefully when performing the Simple ITK resampling step described in the previous section. If there are multiple $z$-spacing values available, we choose the mode of the $z$-spacing across the whole scan as our input to the resampling function. This prevents warping which occurs if the minimum or maximum $z$-spacing value is chosen instead.

### A.2.8 Sanity Checks

Twenty-five CT volumes were selected for detailed evaluation after processing. Based on the abnormality labels automatically extracted from free-text reports, we selected 9 severely abnormal CTs (high count of abnormalities), 9 CTs with zero abnormalities, and 7 random CTs. Each of these CTs was visualized in axial, coronal, and sagittal views, for a total of 75 visualizations to ensure that (a) all slices were stacked in the correct order, (b) no resampling artefacts had been introduced, and (c) the grayscale pixel values appeared reasonable for the different organs. All 25 CT scans passed inspection.

### A.2.9 Efficient Storage of CT Volumes

The Simple ITK resampling step converted the pixel values to 32-bit floats. In the final arrays, we represent pixel values as 16-bit integers. Hounsfield units in DICOMs are all integers and can be represented in 12 bits, but we ultimately use 16 bits because a 12-bit representation is not available in numpy (Van Der Walt et al., 2011). We save the final 3D numpy array using lossless zip compression. The raw unprocessed DICOMs require 9.2 terabytes of storage. The final zipped preprocessed arrays require 2.8 terabytes.

# Appendix B. CT Report Label Extraction

## B.1 Sentence Analysis for Radiology Label Extraction: Term Search

**Table B1 SARLE Term Search.** Medical vocabulary used in SARLE for automatic extraction of 83 abnormality labels from CT reports. The term search is applied only to abnormal phrases. The presence of any word in the "Any" column will result in considering the associated abnormality present. The presence of any word in the "Term 1" column along with any word in the "Term 2" column will result in considering the associated abnormality present. The abnormality will not be considered present if any word in the "Exclude" column is present. "Example Matches" shows example words and phrases that will result in a positive label for that abnormality based on the term search. Entries in all caps represent extensive lists that can be found in the code (*e.g.*, "LUNG_TERMS" which encompasses different phrasings for the lobes of the lung and the right and left lungs.) Note that in our preprocessing all words have been lowercased so we do not have to worry about capitalization.

| Abnormality | Any | Term1 | Term2 | Exclude | Example Matches and Comments |
|---|---|---|---|---|---|
| 'bandlike_or_linear' | 'bandlike', 'band like', 'band-like', 'linear' | | | | |
| 'groundglass' | 'groundglass', 'ground glass', 'ground-glass' | | | | |
| 'honeycombing' | 'honeycomb' | | | | honeycombing |
| 'reticulation' | 'reticula' | | | | reticular, reticulated, reticulation |
| 'tree_in_bud' | 'tree-in-bud', 'tree in bud' | | | | |
| 'airspace_disease' | 'airspace disease', 'copd', 'chronic obstructive' | 'airspace', 'air-space', 'airway', LUNG_TERMS | 'disease' | | airways disease, chronic obstructive pulmonary disease, air-space diseaes |
| 'air_trapping' | 'air trapping' | | | | |
| 'aspiration' | 'aspirat' | | | | |
| 'atelectasis' | 'atelecta' | LUNG_TERMS | 'collapse' | | atelectasis, atelectases, atelectatic, collapsed right upper lobe, collapsed left lower lower, collapse of the right middle lobe |
| 'bronchial_wall_thickening' | 'bronchial wall thicken' | 'bronch' | 'thicken' | | bronchial thickening |
| 'bronchiectasis' | 'bronchiecta' | | | | bronchiectases, bronchiectasis, bronchiectatic |
| 'bronchiolectasis' | 'bronchiolecta' | | | | |
| 'bronchiolitis' | 'bronchiolitis' | | | | |
| 'bronchitis' | 'bronchitis' | | | | |
| 'emphysema | 'emphysem', 'blister', 'bulla', 'bullous' | | | | emphysema, emphysematous, bulla, bullae |
| 'hemothorax' | 'hemothora', 'hemopneumothora' | | | | hemothorax, hemothoraces, hemopneumothoraces |
| 'interstitial_lung_disease' | 'interstitial lung disease', 'interstitial disease', 'interstitial pneumonia', ' uip ', ' ild ', 'fibrosis', ' ipf ', ' nsip ', 'interstitial pneumonitis', 'hypersensitivity pneumonitis', 'organizing pneumonia', 'sarcoidosis' | | | EXCLUDE_NONLUNG | we exclude non-lung terms because we don't want to pick up on liver fibrosis |
| 'lung_resection' | 'pneumonectomy', 'lobectomy', 'bronchial stump' | 'resect' | LUNG_TERMS | | resection, resected, wedge resection of the lower lobe |
| 'mucous_plugging' | 'mucous plug', 'mucus plug' | | | | mucous plug, mucous plugs, mucus plugging |
| 'pleural_effusion' | 'effusion', 'pleural effusion', 'pleural fluid', 'basilar fluid', 'lower lobe fluid', 'fissural fluid' | 'pleura' | 'fluid' | 'pericardial' | pleural effusion; fluid in the pleural space; fissural fluid |
| 'pleural_thickening' | 'pleural thick' | 'pleura' | 'thicken' | | thickened pleura, pleural thickening |
| 'pneumonia' | 'pneumonia', 'pneumoniae' | | | | pneumonia, pneumoniae. do not use the stem 'pneumoni' because that also hits 'pneumonitis' which is different |
| 'pneumonitis' | 'pneumonitis' | | | | |
| 'pneumothorax' | 'pneumothora' | | | | pneumothorax, pneumothoraces, hydropneumothorax |
| 'pulmonary_edema' | 'edema' | | | | |
| 'septal_thickening' | 'septal thickening' | | | | |
| 'tuberculosis' | 'tubercul' | | | | tuberculous, tuberculosis, tuberculoses. nontuberculous excluded with sentence rules for 'non' |
| 'cabg' | ' cabg ', 'bypass' | | | | coronary artery bypass graft/grafting/grafts/surgery, bypass grafting, coronary bypass |
| 'cardiomegaly' | 'cardiomegaly' | 'large', 'increase', 'prominent', ' dilat' | 'cardiac', 'heart', 'ventric', 'atria', 'atrium' | | ventricular enlargement; the heart is enlarged; atrial dilation; increased heart size |
| 'coronary_artery_disease' | 'coronary artery', 'coronary arterial' | 'coronary' | 'disease', 'calci', 'atheroscl' | | coronary artery disease/calcium. if coronary arteries are mentioned in an abnormal sentence, they're abnormal |
| 'heart_failure' | 'heart failure' | 'failure' | 'cardiac', 'heart', 'ventric', 'atria', 'atrium' | | |
| 'heart_valve_replacement' | 'valve replacement' | 'aortic', 'mitral', 'tricupsid', 'pulmonary', 'bicuspid', 'pulmonic' | 'replacement', 'prosthe', 'replaced' | | mitral prosthesis, prosthetic aortic valve, replaced bicuspid valve |
| 'pacemaker_or_defib' | 'pacemaker', ' pacer ', 'pacing device', 'leads', ' icd ', 'defibr' | | | | defibrillator |
| 'pericardial_effusion' | 'pericardial effusion', 'pericardial fluid' | 'pericardi' | 'fluid', 'effusion' | | pericardial effusion present; effusion in the pericardial sac; fluid also seen in the pericardial space |
| 'pericardial_thickening' | 'pericardial thicken' | 'pericardi' | 'thicken' | | |
| 'sternotomy' | 'sternotomy' | | | | |
| 'arthritis' | 'arthritis', 'arthritic', 'degenerative' | | | | |
| 'atherosclerosis' | 'atheroscler' | | | | atherosclerosis, atherosclerotic |
| 'aneurysm' | 'aneurysm' | | | | aneurysm, aneurysmal |
| 'breast_implant' | | 'implant', 'prosthesis', 'prostheses' | 'breast' | | breast implant, breast prosthesis, breast prostheses |
| 'breast_surgery' | 'mastectomy', 'lumpectomy' | | | | |
| 'calcification' | 'calcifi', 'calcium' | | | | calcification, calcified. deliberately excludes 'calculus'. noncalcified excluded with sentence rules for 'non' |
| 'cancer' | 'cancer', 'metasta', 'tumor', 'malignan', 'carcinoma', 'neoplas', 'sarcoma', 'blastoma', 'cytoma', 'melanoma', 'lymphoma', 'mesothelioma', 'myeloma', 'mycetoma' | | | | cancer, cancerous, metastasis, metastases, metastatic, malignancy, malignant, carcinoma, carcinomatous, carcinomatosis, neoplasm, neoplastic, myxofibrosarcoma, liposarcoma, neuroblastoma, plasmocytoma |
| 'catheter_or_port' | 'catheter', ' cath ', 'picc', 'venous line', ' port ' | | | | port needs spaces around it so it doesn't catch 'portion'<br>don't use 'tip' because that will get the tips of gj tubes etc. |
| 'cavitation' | 'cavitation', 'cavitary', 'cavity' | | | | |
| 'clip' | 'clip' | | | | |
| 'congestion' | 'congest' | | | | |
| 'consolidation' | 'consolid' | | | | consolidation, consolidative |
| 'cyst' | ' cyst ', ' cysts ', ' cystic ' | | | 'cystic fibrosis' | cyst, cystic |
| 'debris' | 'debris' | | | | esp. in airways |
| 'deformity' | 'deform' | | | | deformity, deformed, deformation |
| 'density' | 'density', 'densities' | | | | |
| 'dilation_or_ectasia' | ' dilat', 'ectasia', ' ectatic ' | | | | dilation, dilated, dilatation. need spaces around ectatic to avoid confusion with atelectatic. |
| 'distention' | 'disten' | | | | distended stomach, distention of the colon |
| 'fibrosis' | 'fibrosis', 'fibrotic', 'fibroses' | | | | |
| 'fracture' | 'fracture' | | | | fracture, fractures, fractured |
| 'granuloma' | 'granuloma' | | | | granuloma, granulomatous. not to be confused with 'granulation' which is different |
| 'hardware' | 'hardware' | | | | spinal hardware |
| 'hernia' | 'hernia' | | | | |
| 'infection' | 'infect' | | | | infection, infected, infective. noninfectious excluded with sentence rules for 'non' |
| 'infiltrate' | 'infiltrat' | | | | infiltrate, infiltrates, infiltration |
| 'inflammation' | 'inflam' | | | | inflammation, inflammatory, inflamed |
| 'lesion' | 'lesion' | | | | lesion, lesions |
| 'lucency' | 'lucency', 'lucencies' | | | | |
| 'lymphadenopathy' | 'adenopathy'<br>*function* | *function* | *function* | *function* | lymphadenopathy, adenopathy. note: there is a special function that also captures lymphadenopathy via measurements, *e.g.* '2.5 cm lymph node' is lymphadenopathy but '3 mm lymph node' is not. |
| 'mass' | 'mass' | | | | mass, masses |
| 'nodule' | 'nodul' | | | | nodule, nodular, nodularity |
| 'nodulegr1cm' | *function* | *function* | *function* | *function* | note: there is a special function that computes this label based on measurements. |
| 'opacity' | 'opaci' | | | | opacity, opacities, opacification |
| 'plaque' | 'plaque' | | | | |
| 'postsurgical' | 'surgical', 'status post', 'surgery', 'postoperative', 'post operative' | | | | postsurgical findings/changes, 'post surgical', prior surgery |
| 'scarring' | 'scar' | | | | scar, scarring, scarred |
| 'scattered_calc' | | 'scatter' | 'calcifi' | | scattered calcifications |
| 'scattered_nod' | | 'scatter' | 'nodul', 'node' | | scattered nodules, scattered nodes |
| 'secretion' | 'secretion', 'secrete' | | | | esp. in airways |
| 'soft_tissue' | 'soft tissue' | | | | 'soft tissue in the mediastinum' |
| 'staple' | 'staple', 'stapling' | | | | staple, stapled, staples |
| 'stent' | ' stent' | | | | need space in front of stent to distinguish it from 'consistent' |
| 'suture' | 'suture' | | | | |
| 'transplant' | 'transplant' | | | | lung transplant, heart transplant, liver transplant |
| 'chest_tube' | 'chest tube' | | | | |
| 'tracheal_tube' | 'tracheal tube', 'tracheostomy tube' | | | | |
| 'gi_tube' | 'nasogastric tube', 'ng tube', 'gastrojejunostomy tube', | | | | |

| | 'gastric tube', 'esophageal tube', 'gj tube', 'enteric tube', 'feeding tube', 'gastrostomy tube' | | | | |
|---|---|---|---|---|---|

## B.2 Radiology Label Extraction Related Work

Prior work in radiology label extraction has focused on chest x-rays (Demner-Fushman et al., 2016; Irvin et al., 2019; Wang et al., 2017), head CTs (Banerjee et al., 2017; Zech et al., 2018), and thromboembolic disease in chest CTs (Chapman et al., 2011; Chen et al., 2018; Pham et al., 2014). A literature search did not reveal a label extraction approach developed for unenhanced chest CTs for a wide range of abnormalities. The most generalizable prior work includes rule-based systems focused on negation detection, including NegEx (Chapman et al., 2001), ConText (Chapman et al., 2011), and NegBio (Peng et al., 2018). Our work is inspired by these systems but differs in several ways:

- These systems use abnormality tagging with MetaMap (Aronson and Lang, 2010), whereas we were interested in defining a vocabulary that would be targeted to unenhanced chest CT scans and easily explainable to users (the vocabulary is shown above in Table B1). The vocabulary is easy to customize. Users can define finer classes (*e.g.,* distinguishing "dependent atelectasis" from "collapsed lobe") or broader classes (*e.g.,* grouping types of interstitial lung disease together).
- We conceptualize our phrase classification as "abnormality detection" which is a superset of "negation detection" in that "abnormality detection" also includes rules based on words like "patent" which are not negations but do describe a normal state (*e.g.,* "the vessels are patent").
- For the negation detection that we do perform, we take a different approach from NegEx, ConText, and NegBio. NegEx and ConText use predefined "negation scopes" (*e.g.,* 5 words away from "absent") to determine which findings are negated, which we view as a limitation for CT reports due to their variable sentence length and structure. NegBio uses NLTK, the Bllip parser, and Stanford CoreNLP to compute a universal dependency graph for each report; it requires that the abnormality extraction rules are defined over the sentence graph rather than the sentence text directly (Peng et al., 2018). We found that an approach of intermediate complexity was highly effective for extracting abnormalities from our chest CT reports, in which negation scopes are defined directly on the sentence text, but include a direction (forward/backward) and can be limited by other words (*e.g.,* "and", "with") or the beginning/end of a sentence. The entirety of our "abnormality detection" including all negation detection requires fewer than 300 lines of Python code and does not have any dependencies on pretrained models.
- Finally, we define special rules for nodules > 1 cm and lymphadenopathy (lymph nodes > 1 cm). These special rules extract measurements from the sentence text (*e.g.,* "3 mm", "1.2 cm") to determine whether the measurement-sensitive abnormality is present.

**Table B2. Survey of work related to label extraction from radiology reports.** "Ref" denotes the reference by first author and year. "Type" refers to the method type and can be rule-based, report classifier, or hybrid. "#" refers to the number of disease-level labels included in the study. Where available, the list of disease-level labels is provided below the table. "U" indicates whether ambiguity or uncertainty detection is used in any form (Y for yes, N for no).

| Ref | Type | # | U | Dataset | Method Summary | Performance |
|---|---|---|---|---|---|---|
| *Demner-Fushman, 2016*(Demner-Fushman et al., 2016) | Rule-based | >50 | N | Public: OpenI | Term search using the Medical Text Indexer (MTI) followed by negation detection using the Neg-Ex6 algorithm implemented in MetaMap25 software. | Precision 0.54 – 1.0 Recall 0.015 – 0.963 F-Score 0.029 – 0.974 (Calculated in our work, for top 22 most common labels) |
| *Wang 2017\**(Wang et al., 2017) | Rule-based | 9 | N | Public: OpenI | ChestX-ray8. Disease concept mining (term search) using DNorm (SNOMED-CT) and MetaMap (UMLS Metathesaurus). Negation detection with handcrafted rules. Processing with NLTK, Bllip parser, and the Stanford dependencies converter. | Precision 0.66 – 1.0 Recall 0.40 – 0.99 F-Score 0.52 – 0.93 (Table 1) |
| *Peng 2018\**(Peng et al., 2018) | Rule-based | 14 | Y | Public: OpenI BioScope (not radiology specific), PK (not radiology specific) Private: ChestX-ray | NegBio. Medical findings recognition (term search) using MetaMap (UMLS concepts) for 14 disease finding types. Construction of universal dependency graph. Negation and uncertainty detection using rules defined on the universal dependency graph. Processing with NLTK, Bllip parser, and the Stanford dependencies converter. | OpenI Precision: 0.898 Recall: 0.850 F-Score: 0.873 ChestX-ray Precision: 0.944 Recall: 0.944 F-Score: 0.944 (Table 3, MetaMap+NegBio) |
| *Irvin 2019*(Irvin et al., 2019) | Rule-based | 14 | Y | Private: 1000 reports. 2 board-certified radiologists had labels extracted the reports to label whether each observation was present, | CheXpert. Mention extracting (term search) using a large list of phrases manually curated by board-certified radiologists. Mention classification as negative, uncertain, or positive using rule-based pre-negation uncertainty, | F-Score Mention: 0.769 – 1.0 F-Score Negation: 0.720 – 1.0 F-Score Uncertain: 0.286 – 0.936 (Table 1) |

| | | | | | absent, uncertain, or not mentioned | negation, and post-negation uncertainty. Processing with NLTK, Bllip parser, and the Stanford dependencies converter CoreNLP. | |
|---|---|---|---|---|---|---|---|
| *Pham 2014*(Pham et al., 2014) | Report classifier | 3 | N | | Private: 573 radiology reports written in French | A lexicon of 1242 terms was compiled from French MeSH terms, Medcode, and other sources. The Brat interface was used to revise corpus annotation. The ratio of positive to negative cases was manually adjusted. Classification of reports by diagnosis was performed using Waikato Environment for Knowledge Analysis (WEKA), and Wapiti, with a naïve Bayes classifier and a maximum entropy classifier. | Naïve Bayes: Precision: 0.67 – 0.99 Recall: 0.5 – 0.97 F-Score: 0.57 – 0.98  Maximum entropy: Precision: 1.00 – 1.00 Recall: 0.95 – 1.00 F-Score: 0.98 – 1.00 (Table 4) |
| *Zech 2018*(Zech et al., 2018) | Report classifier | 55 | N | | Private: 1,004 labeled head CT reports and 95,299 unlabeled head CT reports | Features were created using bag of words, word embeddings, and Latent Dirichlet allocation–based approaches. The classifier was lasso logistic regression. Processing: stop words and particular phrases were removed; stemming using the Porter stemming algorithm; construction of n-grams | Precision: not reported Recall: 0.903 F-Score: 0.671 (Table 3, all labels) |
| *Chen 2018*(Chen et al., 2018) | Report classifier | 3 | N | | Private: internal contrast-enhanced chest CT reports (final training set size 2,500), and external private data from U. Pittsburgh | CNN with human in the loop. Three iterations: select random sample of 500 reports the CNN labeled as PE positive; radiologist annotates these reports; radiologist's labels treated as truth in the next round of training. | Precision: not reported Recall: 0.950 F-Score: 0.938 (Table pg. 5, CNN internal validation) |

| Banerjee 2018 (Banerjee et al., 2017) | Hybrid | 5 | N | Private: 10,000 labels extracted CT head imaging reports and 1,188 had labels extracted CT head imaging reports | Classifiers: random forest, KNN (n=10), KNN (n=5), SVM (radial kernel), SVM (polynomial kernel). Report representation using word2vec. Processing: NLTK to discard stop words, concatenation of common bi-grams, encoding of negation dependency, common terms mapping using CLEVER terminology, domain-specific dictionary mapping using RadLex and SPARQL | Precision: 0.633 – 0.886 Recall: 0.795 – 0.904 F-Score: 0.704 – 0.891 (Table 4, with domain-specific dictionary) |
|---|---|---|---|---|---|---|

Labels:
- *Wang 2017*: atelectasis, cardiomegaly, effusion, infiltration, mass, nodule, normal, pneumonia, pneumothorax
- *Peng 2018*: atelectasis, cardiomegaly, consolidation, edema, effusion, emphysema, fibrosis, hernia, infiltration, mass, nodule, pleural thickening, pneumonia, pneumothorax
- *Irvin 2019*: no finding, enlarged cardiomegaly, cardiomegaly, lung lesion, lung opacity, edema, consolidation, pneumonia, atelectasis, pneumothorax, pleural effusion, pleural other, fracture, support devices (Table A1)
- *Pham 2014*: pulmonary embolism, deep vein thrombosis, incidentaloma
- *Chen 2018*: pulmonary embolism presence, chronicity, and location
- *Banerjee 2018*: intracranial hemorrhage on a scale of 1 – 5

Acquiring report-level annotation needed to train and evaluate report-level classifiers is labor-intensive:
Zech et al.(Zech et al., 2018) produced 1,004 reports × 55 labels = 55,220 labels total
Banerjee et al.(Banerjee et al., 2017) produced 1,188 reports × 5 labels = 5,940 labels total
Chen et al.(Chen et al., 2018) produced 2,512 reports × 3 labels = 7,536 labels total.

# Appendix C. Whole CT Volume Multilabel Classification

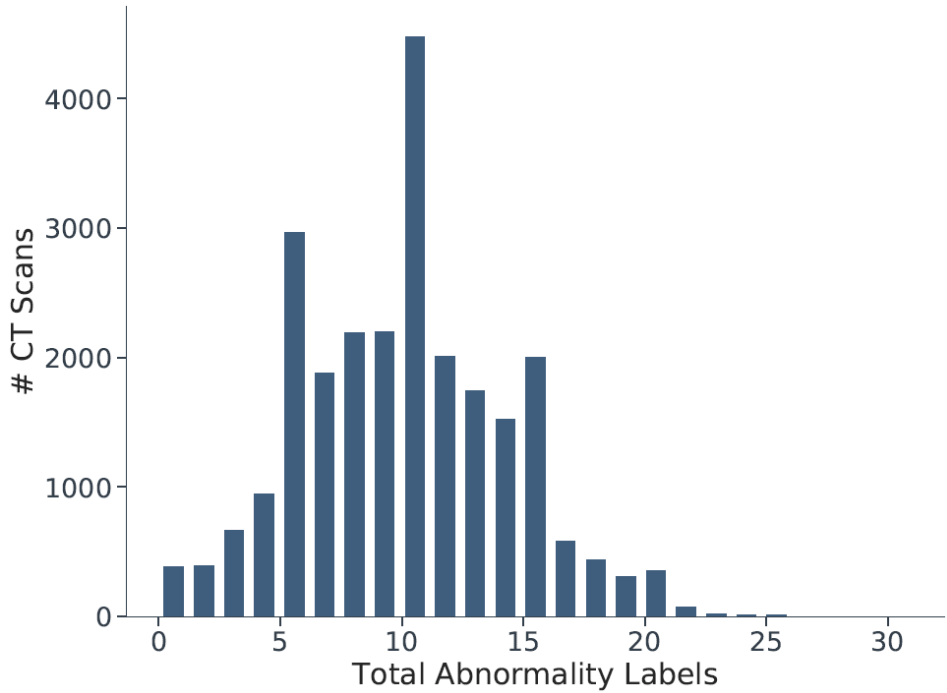

**Figure C1. Histogram of the total count of abnormality labels per CT scan.** Each vertical increment represents a single CT scan. The x-axis represents the counts of abnormality labels in one scan. The median number of abnormalities in one scan is 10, with an interquartile range of 6.

**Table C1. CT-Net-83 Test Set Performance on 83 Abnormalities.**
Test set AUROC and average precision of the CT-Net-83 model across all 83 abnormalities. "CABG" stands for coronary artery bypass graft. "GI tube" includes NG tubes, GJ tubes, and other gastrointestinal tubes. See Appendix B for the medical synonyms included in each abnormality.

| Abnormality | AUROC | Average Precision | Test Set Percent | Test Set Count |
|---|---|---|---|---|
| pacemaker or defibrillator | 0.975 | 0.699 | 0.039 | 279 |
| honeycombing | 0.972 | 0.463 | 0.027 | 193 |
| tracheal tube | 0.971 | 0.597 | 0.017 | 121 |
| lung resection | 0.967 | 0.876 | 0.194 | 1,398 |
| sternotomy | 0.965 | 0.598 | 0.071 | 514 |
| CABG | 0.965 | 0.527 | 0.040 | 288 |
| transplant | 0.963 | 0.751 | 0.057 | 414 |
| catheter or port | 0.954 | 0.716 | 0.105 | 755 |
| heart failure | 0.952 | 0.040 | 0.002 | 18 |
| pleural effusion | 0.951 | 0.869 | 0.195 | 1,404 |
| heart valve replacement | 0.949 | 0.219 | 0.018 | 133 |
| chest tube | 0.944 | 0.387 | 0.020 | 146 |
| GI tube | 0.940 | 0.492 | 0.022 | 162 |
| emphysema | 0.929 | 0.843 | 0.243 | 1,754 |
| pulmonary edema | 0.921 | 0.524 | 0.052 | 373 |
| fibrosis | 0.910 | 0.662 | 0.112 | 811 |
| interstitial lung disease | 0.906 | 0.764 | 0.153 | 1,102 |
| pneumothorax | 0.904 | 0.355 | 0.028 | 205 |
| postsurgical | 0.896 | 0.853 | 0.428 | 3,089 |
| hemothorax | 0.890 | 0.038 | 0.004 | 28 |
| coronary artery disease | 0.873 | 0.830 | 0.494 | 3,563 |
| congestion | 0.873 | 0.018 | 0.002 | 18 |
| bronchiolectasis | 0.871 | 0.120 | 0.016 | 118 |
| pneumonitis | 0.865 | 0.168 | 0.026 | 184 |
| cardiomegaly | 0.851 | 0.462 | 0.090 | 649 |
| reticulation | 0.844 | 0.391 | 0.094 | 681 |
| tree in bud | 0.835 | 0.253 | 0.030 | 217 |
| bronchiectasis | 0.834 | 0.525 | 0.140 | 1,009 |
| septal thickening | 0.831 | 0.287 | 0.071 | 515 |
| tuberculosis | 0.828 | 0.142 | 0.014 | 104 |
| consolidation | 0.816 | 0.427 | 0.151 | 1,086 |
| pneumonia | 0.816 | 0.152 | 0.039 | 282 |
| air trapping | 0.812 | 0.129 | 0.033 | 235 |
| bronchitis | 0.806 | 0.034 | 0.007 | 48 |
| calcification | 0.804 | 0.877 | 0.670 | 4,831 |
| suture | 0.787 | 0.067 | 0.023 | 163 |
| cavitation | 0.786 | 0.143 | 0.035 | 252 |
| breast surgery | 0.781 | 0.125 | 0.018 | 131 |
| clip | 0.779 | 0.238 | 0.084 | 609 |
| staple | 0.778 | 0.060 | 0.020 | 147 |
| mass | 0.773 | 0.392 | 0.120 | 863 |
| aspiration | 0.767 | 0.183 | 0.066 | 474 |
| atelectasis | 0.765 | 0.598 | 0.283 | 2,037 |
| bronchiolitis | 0.765 | 0.083 | 0.018 | 131 |
| secretion | 0.753 | 0.047 | 0.015 | 110 |
| opacity | 0.740 | 0.757 | 0.538 | 3,877 |
| cancer | 0.739 | 0.549 | 0.280 | 2,016 |
| nodule >1cm | 0.730 | 0.303 | 0.118 | 853 |
| lymphadenopathy | 0.728 | 0.372 | 0.167 | 1,203 |
| groundglass | 0.727 | 0.587 | 0.332 | 2,390 |
| breast implant | 0.725 | 0.093 | 0.012 | 86 |
| nodule | 0.718 | 0.883 | 0.779 | 5,617 |
| debris | 0.713 | 0.094 | 0.031 | 223 |
| atherosclerosis | 0.710 | 0.429 | 0.284 | 2,047 |
| plaque | 0.709 | 0.058 | 0.017 | 122 |
| stent | 0.709 | 0.069 | 0.037 | 267 |
| pleural thickening | 0.706 | 0.182 | 0.086 | 621 |
| airspace disease | 0.703 | 0.268 | 0.134 | 965 |
| mucous plugging | 0.701 | 0.125 | 0.037 | 267 |
| infection | 0.697 | 0.557 | 0.333 | 2,398 |
| pericardial effusion | 0.697 | 0.305 | 0.150 | 1,078 |
| bronchial wall thickening | 0.691 | 0.173 | 0.085 | 615 |
| aneurysm | 0.678 | 0.027 | 0.014 | 102 |
| scarring | 0.672 | 0.314 | 0.205 | 1,477 |
| soft tissue | 0.659 | 0.221 | 0.138 | 997 |
| bandlike or linear | 0.654 | 0.242 | 0.155 | 1,115 |
| scattered calcifications | 0.644 | 0.239 | 0.168 | 1,210 |
| deformity | 0.638 | 0.094 | 0.062 | 449 |
| arthritis | 0.635 | 0.376 | 0.285 | 2,053 |
| inflammation | 0.631 | 0.124 | 0.082 | 594 |
| dilation or ectasia | 0.620 | 0.073 | 0.051 | 370 |
| hernia | 0.615 | 0.171 | 0.120 | 866 |
| pericardial thickening | 0.615 | 0.063 | 0.035 | 254 |
| lesion | 0.613 | 0.330 | 0.238 | 1,714 |
| lucency | 0.604 | 0.030 | 0.018 | 133 |
| distention | 0.604 | 0.019 | 0.012 | 86 |
| fracture | 0.600 | 0.087 | 0.065 | 472 |
| cyst | 0.594 | 0.184 | 0.143 | 1,032 |
| granuloma | 0.588 | 0.116 | 0.083 | 595 |
| hardware | 0.577 | 0.020 | 0.017 | 120 |
| density | 0.560 | 0.115 | 0.090 | 647 |
| scattered nodules/nodes | 0.559 | 0.249 | 0.210 | 1,512 |
| infiltrate | 0.526 | 0.016 | 0.015 | 107 |

## *Appendices Additional References*